\documentclass{aa} 
\usepackage{txfonts} 
\usepackage{graphicx}
\usepackage[breaklinks]{hyperref}
\begin{document}
\title {The  $\ion{He}{ii}$ Fowler lines and the $\ion{O}{iii}$ and
$\ion{N}{iii}$ Bowen fluorescence lines in the  symbiotic nova RR Tel}
\author{P. Selvelli  \inst{1} \and  J. Danziger \inst{1} \and P.
Bonifacio  \inst{2,3,1}}
\offprints{P. Selvelli}
\institute {INAF-Osservatorio Astronomico di Trieste, Via Tiepolo 11 -
Trieste,
I-34143 Trieste, Italy \and CIFIST Marie Curie Excellence Team \and
Observatoire de Paris, GEPI, 5, Place Jules Janssen, 92195 Meudon,
France }

\date{Received.....; accepted } 

\abstract 
{} {A detailed study of the $\ion{O}{iii}$ and $\ion{N}{iii}$ Bowen
lines in the spectrum of RR Tel} 
{Absolute intensities for the $\ion{He}{ii}$, $\ion{O}{iii}$,
and $\ion{N}{iii}$ emission
lines have been obtained from STIS, UVES, FEROS and IUE data, after
re-calibration of UVES and FEROS on the STIS scale.} {A new measure of
reddening
(E$_{(B-V)}$$\sim$0.00) has been obtained from the comparison between
the observed and the theoretical intensity decrement for 20 emission
lines of the $\ion{He}{ii}$ Fowler (n$\to$3) series. This value
has been confirmed by the STIS and IUE continuum distribution, and by
the value of n$_H$ from the damped profile of the IS  H Ly-$\alpha$ line.
We have obtained very accurate measurements for about thirty Bowen
lines of $\ion{O}{iii}$ and a precise determination of the efficiency in
the O1 and O3 excitation channels (18 $\%$ and 0.7 $\%$, respectively).
The
relative $\ion{O}{iii}$ intensities are in good agreement with the
predictions by Froese Fischer (1994). A detailed study of the decays from
all
levels involved in the Bowen mechanism has lead to the detection of two
new
$\ion{O}{iii}$ Bowen lines near $\lambda$ 2190. High resolution IUE data
have
shown a nearly linear decline with time, from 1978 to 1995, in the
efficiency
of the O1 and O3 processes, with a steeper slope for the O3 channel. A
detailed
study of the $\ion{N}{iii}$ $\lambda$ 4640 lines and of their excitation
mechanism has shown that, recombination and continuum fluorescence being
ruled
out, line fluorescence remains the only viable mechanism to pump the 
3d $^2D_{5/2}$ and 3d $^2D_{3/2}$ levels of $\ion{N}{iii}$. We point out
the
important role of multiple scattering in the resonance lines of
 $\ion{O}{iii}$ and $\ion{N}{iii}$ near $\lambda$ 374 and show that 
the observed  $\ion{N}{iii}$
line ratios and intensities can be explained in terms of line
fluorescence by
the three resonance lines of $\ion{O}{iii}$ at $\lambda$$\lambda$
374.432,
374.162 and 374.073  under optically thick conditions.} {}


\titlerunning{The $\ion{He}{ii}$ Fowler lines and the $\ion{O}{iii}$ and
$\ion{N}{iii}$ Bowen lines in
RR Tel}

\maketitle

%

\section{Introduction}

RR Tel is one of the slowest of all novae and  is still evolving
in
the nebular stage more than 60 years after the outburst  of Oct. 1944.
RR Tel has  been classified also as  a
symbiotic nova
on account of the presence of an M5 III star in the binary system. 
In 1949 the nova remnant started to shrink back to white dwarf
dimensions: the visual brightness began a  gradual decline, although the
stellar
temperature increased, and the nebular emission showed a slow evolution
toward an increasing degree of ionization (Mayall 1949, Thackeray 1977, 
Murset and Nussbaumer 1993).
A distinctive peculiarity of
RR Tel is the richness and complexity of its  emission spectrum that
spans
a wide range of ionization and excitation. The
simultaneous presence of  spectral features attributable to:~ 1) the hot
nova remnant with T$_{eff}$ about 150,000 K (Hayes and
Nussbaumer 1986, hereinafter HN86),~ 2) the M5 III
companion star with semi-regular Mira-like pulsations (P $\sim$ 387$^d$)
(Henize and McLaughlin 1951,   Feast et al. 1977), ~   3) a
slowly expanding nebula illuminated by the hot remnant,~ 4) the "shocked"
region where, allegedly,  the wind from the giant and the wind of the
nova interact   (Contini and Formiggini, 1999), ~
5) an IR excess whose origin is quite uncertain  (free-free, dust), 
makes    RR Tel an ideal laboratory  for studies  of 
low-density astrophysical  plasmas. 
  
The recent availability of very good quality UVES and STIS data of RR
Tel
has prompted us
to combine them together, with the purpose of fully exploiting the best
characteristics of these two instruments, i.e. the  high
spectral resolution of  STIS in the UV  and  UVES in the optical,  
together with  the wide 
wavelength coverage with absolutely calibrated data by STIS. This has
allowed the detection of weak spectral features and accurate
measurement of absolute emission line intensities (ELI) and  line ratios.
Data from FEROS (with resolution comparable to that of UVES) have been
also used to complement the STIS data for  $\ion{N}{iii}$  in the two
spectral regions near $\lambda$ 4100  and $\lambda$ 4640, not covered
by UVES.   
High resolution  IUE data in the LW region  have also been used to
follow  the changes in the  $\ion{He}{ii}$ and $\ion{O}{iii}$ line
intensities from 1978 to
1995.

\section{The spectroscopic data }

\subsection{HST-STIS data}
We have retrieved from the HST archive the STIS spectra secured on Oct.
10, 2000 as part of program 8098 (see Keenan et al. 2002).  The
instrument setup included the echelle
gratings  E140M and E230M  with  FUV  and NUV  MAMA detectors, and the 
first-order  gratings  G430M and G750M  with CCD detectors.  The
datasets are  listed in Table 1, together  with some relevant   
spectroscopic information.  We note that the  E140 M and E230 M 
echelle gratings 
 together cover the spectral range from 1150 to 3110 \AA,   while the 
 data obtained with the G430M and G750M first-order gratings  cover
the
range from 3025 to 7100 \AA.  The absolute  fluxes are  accurate  
within 8  $\%$  (MAMA) and 5 $\%$  (CCD),  while the wavelengths
are
accurate to 1.5-5.0 km s$^{-1}$  (MAMA) and   3-15  km s$^{-1}$ (CCD) 
(We
refer to  the
www.stsci.edu/hst/stis/performance/accuracy/document.html
page for more detailed   information).

\begin{table}
\caption{The STIS Dataset    }
\begin{center}
\renewcommand{\tabcolsep}{0.1cm}
\begin{tabular}{lllll}
\hline
Dataset   &   Filter-gra  & Resolution & Detector &  Aperture\\
 \hline     
O5EH01010    &     E140M    &  45800    &  FUV-MAMA   & 0.2 x 0.2 \\ 
O5EH01020  &       G430M    &  5330-10270    &  CCD  & 52 x 0.2 \\
O5EH01030   &      E230M    &     30000    &  NUV-MAMA  & 0.2 x 0.2\\
O5EH01040   &     G430M    &  5330-10270    &  CCD   & 52 x 0.2\\
 O5EH01050   &      G430M   &  5330-10270    &  CCD  & 52 x 0.2 \\                                                        
O5EH01060   &     G430M    &  5330-10270    &  CCD   & 52 x 0.2\\
O5EH01070   &     G430M    &  5330-10270    &  CCD   & 52 x 0.2\\
O5EH01080   &     G430M    &  5330-10270    &  CCD   & 52 x 0.2\\
O5EH01090   &       E230M   &  30000    & NUV-MAMA  & 0.2 x 0.2\\
O5EH010A0   &     G430M    &  5330-10270    &  CCD & 52 x 0.2 \\
O5EH010B0   &     G430M    &  5330-10270    &  CCD & 52 x 0.2 \\
O5EH010C0   &     G430M    &  5330-10270    &  CCD & 52 x 0.2 \\
O5EH010D0   &     G430M    &  5330-10270    &  CCD  & 52 x 0.2\\
O5EH010E0   &     G750M    & 4870-9950    &  CCD  & 52 x 0.2\\
O5EH010F0   &     G750M    & 4870-9950    &  CCD & 52 x 0.2 \\
O5EH010G0   &     G750M    & 4870-9950    &  CCD  & 52 x 0.2\\
\hline
\end{tabular}
\end{center}
\end{table}

 The STIS data have been
recently (Dec. 2003) re-calibrated to allow for the effects of
increasing  CCD
charge
transfer inefficiency  (due to the accumulated effects of radiation
exposure) and for the effects of time-dependent optical
sensitivity. The work in the present  paper is based on these recent
data.

\subsection{UVES data}

For a description of the UVES spectra (obtained on Oct. 16th, 1999) and 
their reduction we refer to Selvelli and Bonifacio (2000). We  
note that the spectral  resolution on UVES  is close to 55000, 
for the red arm spectra, taken with a 0\farcs{8} slit and slightly
higher ($\sim$ 65000) for the blue arm spectra, where the slit
was 0\farcs{6}.
The seeing conditions
during
UVES observations (mid Oct. 1999) were around 0.72-0.75 arcsec.  The
angular size  of the RR Tel nebula is less than 0.1 arcsec, since its
radius is near to 1.0$\times$10$^{15}$ cm (HN86) and the distance is
about 
3500 pc (see section 3.5). 

No spectrophotometric standard star was observed that night,  so the
data presented in Selvelli and Bonifacio (2000) were simply
normalized to the continuum.
The instrument response function may be, nevertheless, estimated
by using observations of a spectrophotometric standard star
taken on a different night.
For this we used the observations of the white dwarf
EGGR 21 (Stone \& Baldwin 1983, Bessel 1999) taken 
on September 26th 1999, observed with the same spectrograph setting
as RR Tel but with a slit width of $10''$.
The UVES spectrum thus calibrated is not suitable for flux measurements
due to: a) the unknown slit losses, when observing with a narrow slit
(smaller than the seeing FWHM); b) possible differences in 
atmospheric extinctions between the night in which RR Tel was
observed  and the night in which EGGR 21 was observed.
We therefore decided to make use of the flux calibrated
STIS spectrum to re-calibrate the UVES spectrum,
in the ranges  $\lambda$$\lambda$3040-3820
and $\lambda$$\lambda$ 5500-5700.
We did so by forcing the continuum of the UVES spectrum
to coincide with that of the STIS  absolute  distribution after proper
f$_{\lambda}$  correction in selected
line-free windows.

This has allowed us to obtain  reliable emission line intensities (ELI)
and reliable line ratios
also for
some (weak) $\ion{O}{iii}$ emission lines that are clearly observed in
UVES
spectra but are not detected or are confused with noise in the
STIS grating spectra in the optical range.

This procedure could be criticized since the STIS data
have been secured about 1 year after the UVES ones, and the star flux
is not constant with time. However, it is justified by the fact that
the variations in the continuum and ELI of RR Tel
are  supposed to be  quite slow with a time-scale of years (Murset and
Nussbaumer 1993,
Zuccolo et al. 1997).
 It is also justified, "a
posteriori", by
the fact that the ELI measured
in STIS and UVES spectra (after this recalibration) give a nearly
constant ($\sim 1.16$ $\pm$ 0.03) ratio for those $\ion{O}{iii}$ lines
that are
common to the two groups of spectra. See also Section 4.5 for a
descripion of the  absolute ELI changes in
the last twenty years for  $\ion{He}{ii}$ and $\ion{O}{iii}$.

From the STIS/UVES ELI
ratios nothing can
be said about any absolute ELI changes between Oct. 1999 (UVES) and Oct.
2000 (STIS). One can only guess that most ELI have followed the
general trend of a slow decline, following the continuum, with the
possible exception of the high excitation and/or high ionization  lines
whose ELI could have
increased.

\subsection{ IUE data}
 
The IUE spectra have been retrieved from the INES (IUE Newly Extracted
Spectra) final archive. For a detailed description of the IUE-INES
system see Rodriguez-Pascual et al. (1999) and Gonzalez-Riestra et al.
(2001). 

 For a new  determination of the UV 
continuum  (see section 3.2) we
have used all  available low resolution IUE spectra taken with the large
aperture,  with the exception
of the very few spectra (e.g. SWP05836, SWP05885, SWP13730, SWP4603)
that have shown saturation in
the  continuum or other problems in the exposure.

High resolution IUE-INES data have been  used in section 4.5 for
measuring  the
intensity in  1978-1995 of some  $\ion{He}{ii}$ Fowler
lines
($\lambda$  2733 and $\lambda$  3203) and some $\ion{O}{iii}$ Bowen
lines
($\lambda$$\lambda$ 2836,
3047,
 3122,
and  3133). The  LWP and LWR spectra are those reported  in Table
 6, and have been individually checked for their quality.  They are 
 not or  only mildly affected by saturation effects or  by  
camera  artifacts.  
 A few IUE spectra that were  over/under exposed  or  affected by
factors such as bad guiding, high background noise, microphonic noise,
etc.  have been  disregarded.  
The quality flags for the $\lambda$ 3133  line  generally indicate
the
presence of a
fiducial mark near the spectrum, but this reseau is located outside of
it. The quality flags indicates also the marginal presence of a few
saturated pixels in the spectra with longer exposure time.

\subsection{ FEROS data}  

We made use of the FEROS commissioning data (Kaufer et al. 1999)
which are available through the FEROS Spectroscopic Database at the LSW
Heidelberg
(http://www.lsw.uni-heidelberg.de/projects/instrumentation/Feros/ferosDB)
to measure the $\ion{N}{iii}$ lines near  $\lambda$ 4640 and $\lambda$
4100  (see section 6.2), which are not covered by the spectra of the
other instruments.   
These  4 spectra were acquired on October 7th 1998, one has an
exposure time of 300s, the other three of 600s.
The standard extracted, flat-fielded, wavelength calibrated, and merged 
spectrum of both object and sky fibres have   been downloaded
from the database. The spectra cover the range
3520--9210 \AA~ with a resolution  $R\sim 60000$.
Also for FEROS, in a similar way as that for UVES,  we have made  use of
the flux calibrated
STIS spectra to re-calibrate the FEROS spectrum, 
in the range  $\lambda$$\lambda$4100-4700.
We did so by forcing the continuum of  FEROS 
to coincide with that of  STIS in selected
line-free windows.
This has allowed us to measure reliable ELI
and reliable line ratios
for a few
 $\ion{N}{iii}$  emission lines that are clearly observed in FEROS
spectra but are not detected or are confused with noise in the
STIS grating spectra.
The same warnings and considerations given in section 2.2 are valid 
here. However, also in this case, the procedure is justified by the
fact that the ELI measured
in STIS and  (re-calibrated) FEROS spectra  give a nearly
constant ($\sim 1.18$ $\pm$ 0.02) ratio for the
$\ion{N}{iii}$ lines that are in 
common to the two spectra.

\section{The reddening and the distance}

\subsection {The $\ion{He}{ii}$ recombination lines }

We have adopted and updated the method
of Penston et al (1983) of comparing the observed intensities of the He
II (n$\to$3) recombination lines with the corresponding theoretical
ratios
in order to obtain an estimate of the color excess E$_{B-V}$.
Incidentally we point
out, that  these (n$\to$3) transitions are often named as "Paschen"
lines in the
astronomical literature, irrespective of their belonging to HI or
$\ion{He}{ii}$,
while the $\ion{He}{ii}$ (n$\to$3) lines should be correctly named as
"Fowler"
lines (as in Table 4.3 of Osterbrock and Ferland 2006) and hereinafter
they will be named so.

The STIS data provide an optimum coverage for the whole set of the
$\ion{He}{ii}$
Fowler lines, from $\lambda$  4685.71 down to the region of the
head
of the series near $\lambda$ 2063  (Fig.1).

\begin{table}
\caption { $\ion{He}{ii}$ Fowler Emission Lines. The air wavelength
(\AA), the ELI 
 (10$^{-13}$erg cm$^{-2}$ s$^{-1}$)
and the FWHM  (km s$^{-1}$) of the $\ion{He}{ii}$ emission lines 
in STIS.  The 
theoretical
values I$_{th}$ (Storey and Hummer, 1995)
correspond to log N$_e$=6.0 and  T$_e$=12,500 K.}
\begin{center}
\renewcommand{\tabcolsep}{0.25cm}
\begin{tabular}{crcrcr}
\hline
$\lambda_{air}$  &  ELI & FWHM & I$_{obs}$/I$_{4686}$ &  
I$_{th}$/I$_{4686}$ & ratio  \\
\hline
4685.71  &147.50& 56.2 &  100.0   &  100.0  &  1.00 \\           
3203.10  &63.52 &54.2  &  43.0     & 43.7  &  0.99   \\                           
2733.30  & 32.33& 56.1   &  21.9 &     23.1  &  0.95   \\                             
2511.20  &18.98 &56.4   &12.9   &  13.9    &0.93  \\                  
2385.40 &12.60& 57.3  &  8.54 &    9.03  & 0.95  \\                    
2306.19  &9.23 &56.2  & 6.26  &   6.25 &  1.00   \\                             
2252.69  &6.25&56.1  &  4.24  &   4.52&   0.94   \\                     
2186.60  &3.96 &63.5  & 2.68   &  2.60&   1.03   \\                         
2165.25  &2.81 &51.4  & 1.91 &     2.01 &  0.95  \\                         
2148.60 & 2.60& 55.1  &   1.76  &   1.68 &  1.05  \\                         
2133.35 & 1.98 &55.1  &  1.34   &  1.40  & 0.96 \\                        
2124.63  &1.80 &50.8  &  1.22  &   1.18 &  1.03 \\                               
2115.82 &  1.51& 48.2  & 1.02    & 1.02  & 1.00  \\                               
2108.50  &1.34 & 56.2  &0.91&     0.90 &  1.01  \\                                
2102.35 & 1.17&52.8   & 0.79  &   0.79&   1.00    \\                            
2097.12 & 1.00&54.8  &  0.68   &  0.71 &  0.96  \\                                  
2092.64  & .94 &55.9  & 0.64   &  0.64  & 1.00   \\                         
2088.72  &.85 & 50.3 &   0.58&     0.59&   0.98   \\                         
2085.41 &   .80&52.5    &0.54  &   0.55 &  0.98   \\                        
2079.88  & .70&53.6   &  0.475  &  0.47&  1.01  \\                                
1640.42 &831.10&63.2  &  563.4  &  702.1&  0.81\\   
\hline
\end{tabular}
\end{center}
\end{table}

The spectral region between $\lambda$$\lambda$ 2025  and  2300 
is
quite interesting 
because it has remained "unexplored" until recent times: all of the IUE
spectra suffered from severe underexposure in this range because of the
extremely low response of the LW cameras below $\lambda$ 2300, and
the HST-GHRS data  cover only  a short range in some selected  spectral
regions. 
 In the
STIS spectra,  the  very good spectral resolution allows one to
 resolve the lines up to the (37$\to$3) transition at $\lambda$  2063.49,
very near to the series limit  (Fig. 1). We recall that in
the UVES spectra the hydrogen Balmer lines were instead resolved up to
the (38$\to$2) transition (Selvelli and Bonifacio, 2000).

\begin{figure}
\centering
\resizebox{\hsize}{!}{\includegraphics{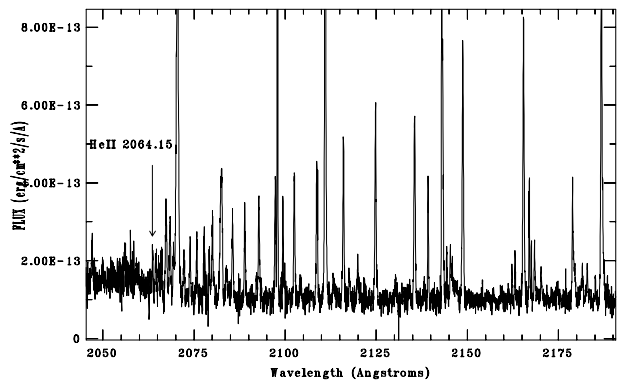}}
\caption{ The  $\ion{He}{ii}$ Fowler lines near the head of the (n-3)
series. 
The last resolved  line is  (37-3) $\lambda$   2063.49.}
\end{figure}

In a study of RR Tel from IUE data, HN86  from
various  diagnostic methods based on 
ratios of collisionally excited lines,  have derived  log  N$_e$
$\sim$ 6.0  and  T$_e$ $\sim$ 12500 K.
Adopting these values (we note, however, that  the relative  lines
strengths in the recombination spectrum of $\ion{He}{ii}$
 have little dependence on electron temperature and density)  we have
compared our observed relative strengths of the various $\ion{He}{ii}$ 
Fowler
lines (normalized to I$_{ 4686}$=100) with the theoretically derived
ratios  for  "case B"  by Storey
and Hummer (1995).  The "case B" assumption is
justified  by the  presence
itself of strong $\ion{O}{iii}$ Bowen lines (that require large optical
depth in
the $\ion{He}{ii}$ Ly-$\alpha$ line, see in the following) and by the
observed
I$_{3203}$/I$_{ 4686}$ ratio
$\sim$ 0.43,  that is generally associated with
"case B" (see Schachter et al., 1991).

Table 2 gives  the observed $\ion{He}{ii}$ ELI (10$^{-13}$erg cm$^{-2}$
s$^{-1}$)
 for all transition of the Fowler
series
 starting from $\lambda$ 4686 (4$\to$3) up to  2079.88 (25$\to$3), with
the
exception of the $\lambda$  2214.67 (11$\to$3) and $\lambda$  2082.47
(24$\to$3) lines 
that are blended.  Table 2  gives also the gaussian FWHM  in km s$^{-1}$ 
(deconvolved from
the instrumental FWHM), the observed and theoretical (for case B log
N$_e$ = 6.0 and T$_e$ = 12,500 K)  intensities relative to
I$_{4686}$=100, 
and, in the last column, the ratio I$_{obs}$/I$_{4686}$ over
I$_{th}$/I$_{4686}$.
In  the last line of Table 2 
the corresponding values for the $\ion{He}{ii}$ Ba-$\alpha$ line at
$\lambda$ 1640
are
also given. The average FWHM
for
20 unblended
lines of the Fowler series is = 53.5 $\pm$ 3.5 km s$^{-1}$. 
In Fig. 2  the I$_{obs}$/I$_{th}$ ratio of Table 2 is plotted versus 
wavelength. The mean value of the
points 
is
0.986, the median is 0.995, with  standard deviation = 0.033. The
points clearly define a
straight line with slope very near 0.00, thus  clearly indicating  that 
E$_{(B-V)}$  $\sim$ 0.00.

We recall that Penston et al obtained E$_{(B-V)}$ = 0.10, but their
measurements
were based on just a few  $\ion{He}{ii}$ lines and in spectra that
were very noisy below $\lambda$ 2400.

In this context, it should be pointed out that under the nebular
conditions
as in the RR Tel nebula, the large optical depth in the $\ion{He}{ii}$
Ly-$\alpha$
line and its complex line transfer that is associated with the Bowen
fluorescence (see Sect. 4) should not influence the $\ion{He}{ii}$
Fowler
lines, but could possibly affect the intensity of the $\ion{He}{ii}$
Ba-$\alpha$
(3-2) line at $\lambda$ 1640.  In  fact,  the  low I$_{1640}$ over
I$_{4686}$
observed ratio = $\sim$ 5.63  (instead of the theoretical one, close to
7.0, for  log N$_e$ = 6.0 and   T$_e$ =12500) is an indication
that
the line has developed some optical depth as a consequence of the
"trapping" of the $\ion{He}{ii}$ Ly-$\alpha$ line $\lambda$ 304  that
causes
level 2 to be
overpopulated.  Also the larger width of the $\lambda$ 1640 line (63 km
s$^{-1}$ as compared to  the 55 km s$^{-1}$  as  the average value for
the remaining $\ion{He}{ii}$ lines) is indicative of its larger optical
depth.

We note that the  low I$_{1640}$ over I$_{ 4686}$  observed 
ratio comes from the 
low intensity of the $\lambda$ 1640 line and not by an excessive
intensity of
 the $\lambda$    4686 line   since 
the
decrements of all of the
Fowler lines relative to I$_{ 4686}$ are very close to the expected
values  (e.g. the I$_{ 3203}$/I$_{ 4686}$ observed ratio is $\sim$
0.43,
to be
compared to the
theoretical ratio $\sim$ 0.44).
It should be also mentioned that Netzer et al. (1985) in a study of
Bowen fluorescence and $\ion{He}{ii}$ lines  in active galaxies and
gaseous
nebulae 
 have considered
the possibility of   an increase in the population  of n=4 of 
$\ion{He}{ii}$ through hydrogen Ly-$\alpha$ pumping from  n=2 to
n=4 of
$\ion{He}{ii}$.
This would result in an enhanced I$_{ 4686}$ (4$\to$3) over I$_{1640}$
(3$\to$2)  ratio. 
However, this mechanism is not effective in RR Tel since, as mentioned, 
the intensity of the various 
Fowler lines relative to I$_{ 4686}$ is very close to the expected one.

\subsection{ The STIS UV + optical continuua }

Penston et al. (1983) confirmed the E$_{(B-V)}$ = 0.10 value obtained
from
the
$\ion{He}{ii}$ lines with the presence (although rather weak) of the
common
interstellar  absorption bump near
$\lambda$ 2200 in the continuum of IUE low resolution spectra.

In order to check for the alleged presence of this absorption bump we
have exploited the high spectral resolution of STIS in the UV region and
plotted the STIS continuum intensity versus wavelength in line-free
regions. Fig. 3 is self explanatory : the very flat near UV continuum
distribution, and the intensity increase towards the far UV, is not
consistent with  what is generally observed in reddened objects.
In Fig. 3 one can easily note the
presence of the two bound-free discontinuities of $\ion{He}{ii}$ and HI
near
$\lambda$ 2060 and $\lambda$ 3646 respectively with D(He II-Fowl)= log
I$_{2060+}$/I$_{2060-}$ $\sim$ 0.21 and  D(HI-Ba) = log
I$_{3646+}$/I$_{3646-}$ $\sim$ 0.48.

\begin{figure}
\centering
\resizebox{\hsize}{!}{\includegraphics[angle=-90]{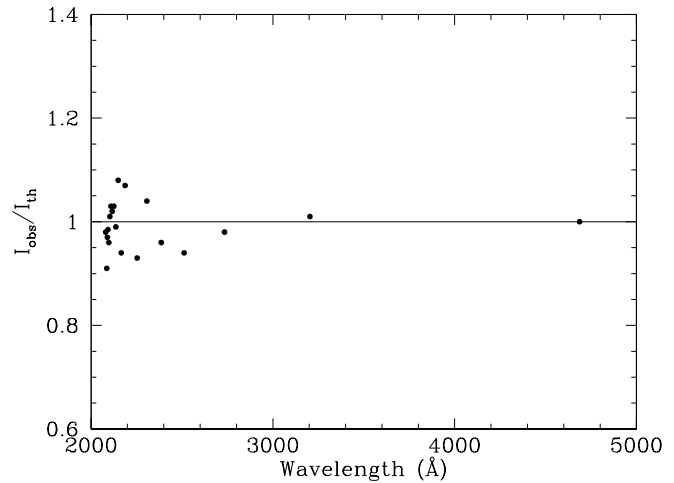}}
\caption{The I$_{obs}$/I$_{th}$ ratio  relative to  I$_{ 4686}$  for the
$\ion{He}{ii}$ Fowler lines  listed  in Table 2.}
\end{figure}

\begin{figure}
\centering
\resizebox{\hsize}{!}{\includegraphics[angle=-90]{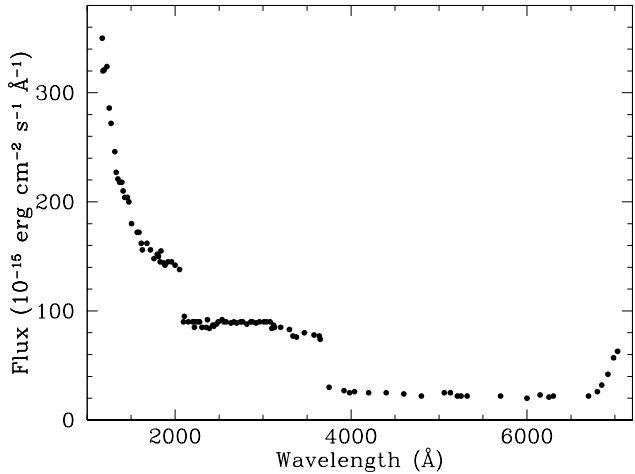}}
\caption{ The observed continuum of RR Tel, determined from line-free
regions in the STIS spectrum. The b-f discontinuities of $\ion{He}{ii}$
(Fowler
series) and
HI (Balmer series) are clearly evident. The rise of the continuum
longward of $\lambda$ 6800  is due to the contribution from the M giant
companion star}
\end{figure}

In order to elucidate further  the origin of the alleged IUE bump
reported by Penston et al.(1983), we
have
coadded and merged all the IUE low resolution spectra available from
the IUE VILSPA archive (large aperture only, not overexposed in the
continuum) and created an "average" spectrum out of 39 SW and 35 LW
spectra (Fig 4). As result of  the good S/N ratio also in the region
below $\lambda$
2300, an inspection of the "average" spectrum definitely shows that
there is no evidence of the IS absorption bump centered near lambda
2175. The absorption dip reported by Penston et al (1983) comes
probably from the "lack" of strong emission features  in the
region near $\lambda$ 2200 (that could have have had the
effect of "raising" the continuum in low resolution IUE spectra), and
from the presence of the Fowler
discontinuity. Also in the emission-line-free region near $\lambda$
2600 the effect is to
mimic the presence of a similar (wide) absorption feature.

A comparison between the IUE continuum and the STIS  continuum also
confirms
the near grey decline with time of 
 the UV continuum.

\begin{figure}
\centering
\includegraphics[width=8.6cm]{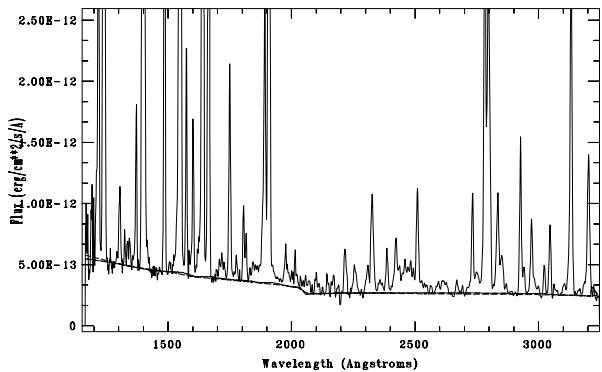}
\caption{ The UV continuum in the average spectrum obtained by  
averaging  and
merging   
39 SWP  and 35 LW  low resolution  IUE spectra. }
\end{figure}

\subsection{The N$_H$ column density towards RR Tel}

From the profile of the damping wings of the IS Ly-$\alpha$ absorption
line (Fig. 5)
one can obtain  the column density  of neutral hydrogen  towards RR Tel.
Only the blue wing can be used, since the red wing is 
affected by the rather strong   $\ion{O}{v}$] emission line  $\lambda$
1218.34.
The blue wing itself is partially contaminated by the   [$\ion{O}{v}$] 
$\lambda$
1213.81
  emission line.
This adds further uncertainity to the derived
column density. 
Our best-fit value, obtained by excluding the interval
contaminated by $\ion{O}{v}$]  $\lambda$ 1213.80, is
N$_H$ $\sim$ 6.9  10$^{19}$ (atoms cm$^{-2}$) 
with a probable relative error of 10\%.   
Strictly speaking  this N$_H$ value 
should be considered as an upper
limit since  a
circumstellar contribution to  N$_H$ cannot be ruled out. 
Jordan et al. (1994) have instead obtained a value for  N$_H$
in the range 1.7-3.1 10$^{20}$  from ROSAT PSPC observations.
The mean
relations between neutral hydrogen column density and dust by
Diplas and Savage (1994) (N$_H$=4.93 10$^{21}$ E$_{(B-V)}$)  gives  
E$_{(B-V)}$=0.014, in good agreement with our estimates.

The HEASARC N$_H$ tool
(heasarc.gsfc.nasa.gov/cgi-bin/Tools/w3nh/w3nh.pl)
 in which
N$_H$   is derived from the HI map by Dickey and Lockman 
(1990) 
indicates a total  galactic
column density N$_H$ in the direction of RR Tel
very close to  4.5 10$^{20}$ (atoms cm$^{-2}$). 
It should be pointed out, however, that  the 
 N$_H$ tool gives the average value for 1 degree x 1 degree bins. 

The valuable Web site "Galactic Dust Extinction
Service" (http://irsa.ipac.caltech.edu/applications/DUST/)
  employs the 100-micron intensity  map of the sky by Schlegel et al
(1998) to provide the foreground (Milky Way) extinction for a line
of
sight  and/or region of the sky.  This tool gives E$_{(B-V)}$=0.05 as
the
total  extinction in the line of sight
of RR Tel. 
The spatial resolution of the map is higher  than for N$_H$  (about 5
arcmin)
but, 
as clearly stated in the cautionary notes,  the various assumptions made
in
the
conversion from 
 the 100-micron dust column density to the color excess  E$_{(B-V)}$ may
affect
the resulting 
 E$_{(B-V)}$ value.

\subsection{Comments on the reddening }

The observations described in the previous sections have convincingly
shown  that   E$_{(B-V)}$  $\sim$ 0. 

We recall that  previous E$_{(B-V)}$ values in the literature  range
from
values as high as  0.7, estimated by Walker (1977) from the  Balmer
decrement,   to  values close to 0.08-0.10,
(Penston
et al., 1983,  Feast et al (1983), Jordan et al., 1994 , Espey et al.,
1996), down to values close to zero (Glass and
Webster, 1973).
The  generally accepted  E$_{(B-V)}$ values  in the literature are
in the range  0.08-0.10. 
Recently, Pereira et al. (1999),  using ESO data with
spectral resolution of $\sim 2.5$ \AA~ (FWHM) have obtained  E$_{(B-V)}$
= 0.04 from the relative ratios of  H$_{\gamma}$/H$_{\beta}$,
H$_{\delta}$/H$_{\beta}$
and $\ion{He}{ii}$ 3203/4686.
Very recently,  Young et al. (2005) have derived E$_{(B-V)}$ $\le$ 0.28   
from  a study of the $\ion{Fe}{vii}$  lines, based on the same STIS and
UVES
spectra of RR  Tel as this work. The authors have pointed out, however,
that the extinction is not well constrained by the $\ion{Fe}{vii}$ lines
and
have derived  E$_{(B-V)}$ $\sim$ 0.11 from the comparison of the
observed and the theoretical (Galavis et al., 1997) NeV 
I$_{2974}$/I$_{1574}$ line ratio. 

We therefore emphasize that the E$_{(B-V)}$  value 
obtained in the present study with the method of the $\ion{He}{ii}$
Fowler lines  
is based on data from 20 lines and is extremely well
constrained, because of the  wide wavelength coverage of the lines and 
because of the fact that most of them  fall  close to  the region near
$\lambda$ 2200 , where the maximum of the extinction occurs.   
Moreover, if, as an exercise,  the observed STIS continuum  plotted in
Fig. 3 were "dereddened" assuming  E$_{(B-V)}$=0.10, the resulting
continuum  would show an artificial emission bump near $\lambda$ 2000.

\begin{figure}
\centering
\resizebox{\hsize}{!}{\includegraphics{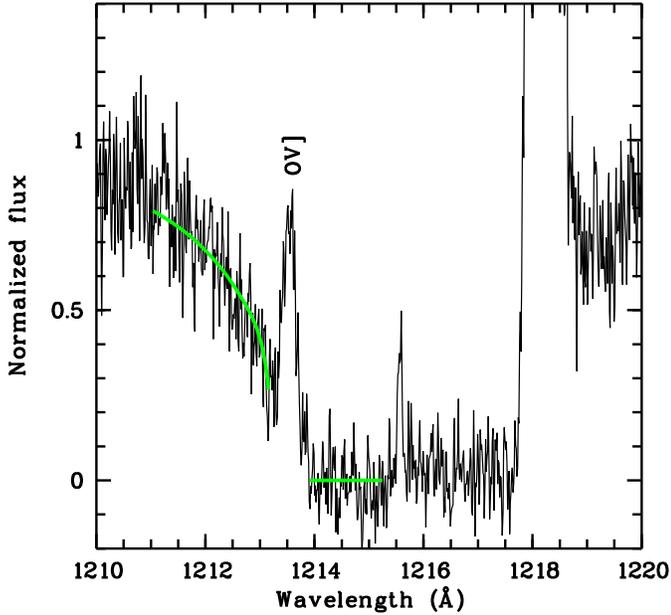}}
\caption{The interstellar Ly$\alpha$ line. Our best fit to the damping
wing, excluding the  [$\ion{O}{v}$] $\lambda$  1213.81 emission line,
implies a
hydrogen column density N(H)=6.9$\times$ 10$^{19}$ cm$^{-2}$
and is shown as a solid line.}
 \end{figure}

\subsection{The distance to RR Tel}

The photometric and spectral development of RR Tel during the OB phases
was that of an extremely slow nova that was near maximum between  the
end
of 1944, when the outburst occurred,  until June 1949 (Friedjung, 1974, 
Heck and Manfroid 1985). It seems
therefore reasonable  to assume that the nova luminosity was
near-Eddington during these decay phases. For a WD mass near 0.6
M$_{\odot}$, as
expected in an extremely slow nova (Livio, 1992),  this assumption gives
M$_{Bol}^{max}$  $\sim$ -6.1, and assuming a bolometric correction  BC
$\sim$
-0.1 for an object with T $\le$ 10000 K,
we obtain M$_v^{max}$ = -6.0, in good agreement with the estimates from
the
various relations (Della Valle and Livio 1995, Downes and Duerbeck
2000) between the absolute magnitude at maximum and the rate of decline
(MMRD).

From the observed m$_v^{max}$=6.7 and our new value for the extinction
(A$_v$=0.0) a distance of 3.47 kpc is obtained.

This value is in good agreement with that of 3.6 kpc obtained by Feast
et al (1983) on the assumption that a Mira is present in RR Tel.
Whitelock (1988) has instead derived d $\sim$ 2.6 kpc by applying a
period-brightness relation to the IR magnitudes and the pulsation
period of the Mira in the system.

From our estimated distance (d = 3.47 kpc) and the total IS hydrogen
column density derived from the Ly-$\alpha$ damped profile (N$_H$=6.9
$\times 10^{19}$ atoms cm$^{-2}$ ) one obtains that the average hydrogen
number
density
N(H) (atoms cm$^{-3}$) is $\sim$  6.7 10$^{-3}$. We recall that Diplas
and Savage (1994) found limits of
0.017-8.62 for the hydrogen number density in the galactic ISM.
Therefore, the intervening ISM towards RR Tel seems  extremely poor in
the content of  dust and neutral hydrogen.

\section{The $\ion{O}{iii}$ Bowen lines} 
\subsection{The $\ion{O}{iii}$ Bowen fluorescence mechanism}

In
high-excitation planetary nebulae and symbiotic stars  a significant
fraction of the EUV - soft X-ray flux (at wavelengths shortward of
the  $\ion{He}{ii}$ Lyman limit, h$\nu$ $\ge$ 54.4 eV)   is absorbed by
the
nebula
and converted into the $\ion{He}{ii}$ Ly-$\alpha$ $\lambda$ 303.782
emission
line 
that may then attain a high intensity.  These  $\lambda$ 303.782 photons
can ionize both H
and
 He and therefore  they  have a strong influence on  the
ionization and
temperature structure of the nebula.

Scattering of these $\ion{He}{ii}$ Ly-$\alpha$  resonance photons  may
result in
one
of the
following processes (See Aller 1984, Osterbrock and Ferland  2006): ~1)
~ escape from the
nebula by diffusion in the wings, ~2)~
photoionization of He$^o$ and H,  ~3)~ absorption by  $\ion{O}{iii}$: in
a
 near coincidence in wavelength with the $\lambda$  303.800 
resonance transition of  $\ion{O}{iii}$ the $\ion{He}{ii}$ Ly-$\alpha$
$\lambda$
303.782 
photons may excite its 2p3d $^3P_2$ level at 40.85 eV (O1 process).

The most likely radiative  process from the excited 2p3d $^3P_2$
level of  $\ion{O}{iii}$
 (total probability about 0.98=0.74+0.24) is that of resonant
scattering, by emission of one of the two resonance lines at lambda
 303.800 (2p3d $^3P_2$ -  2p$^{2}$ $^3P_2$) (O1  line) and lambda 
303.622
(2p3d
$^3P_2$ - 2p$^{2}$
$^3P_1$) (O2  line).

A less probable decay process from the aforesaid 2p3d$^3P_2$  level
(total probability
about 1.85 $\%$) is the cascade through the high lying 2p3p $^3P$
(E $\sim$ 37.23 eV),
2p3p $^3S$ (E $\sim$ 36.89 eV)   and 2p3p $^3D$ (E $\sim$ 36.45 eV)
terms,
by
emission of one of six
subordinate lines (the primary cascade-decays) in the near-UV optical
region of the spectrum (from $\lambda$ 2807.90 to $\lambda$ 3444.06).
However, despite the low probability of the decay
channel through the
subordinate lines, if the optical depth in the $\ion{He}{ii}$
Ly-$\alpha$ line is
very high the trapped $\ion{He}{ii}$ Ly-$\alpha$ photons are repeatedly
scattered and a significant fraction  of
them  can be  converted into the fluorescent
cascade
of  optical and ultraviolet lines.
Unlike
the resonance and near resonance lines that are scattered many times,
the
photons produced by these subordinate 
transitions can  easily escape from the nebula, and, together with the
subsequent cascades (the "secondary" decays), that produce additional
emission lines, (most of them in the optical U spectral range), comprise
the Bowen fluorescence lines   (Bowen, 1934). 

 Fig. 6 represents a partial Grotrian diagram of $\ion{O}{iii}$ that
includes the
most common Bowen O1 transitions. Fig. 6 is complemented by Table 3 that 
   includes the configuration and  the energy (eV and cm$^{-1}$) of the
 relevant levels.

\begin{figure}
\centering
\resizebox{\hsize}{!}{\includegraphics{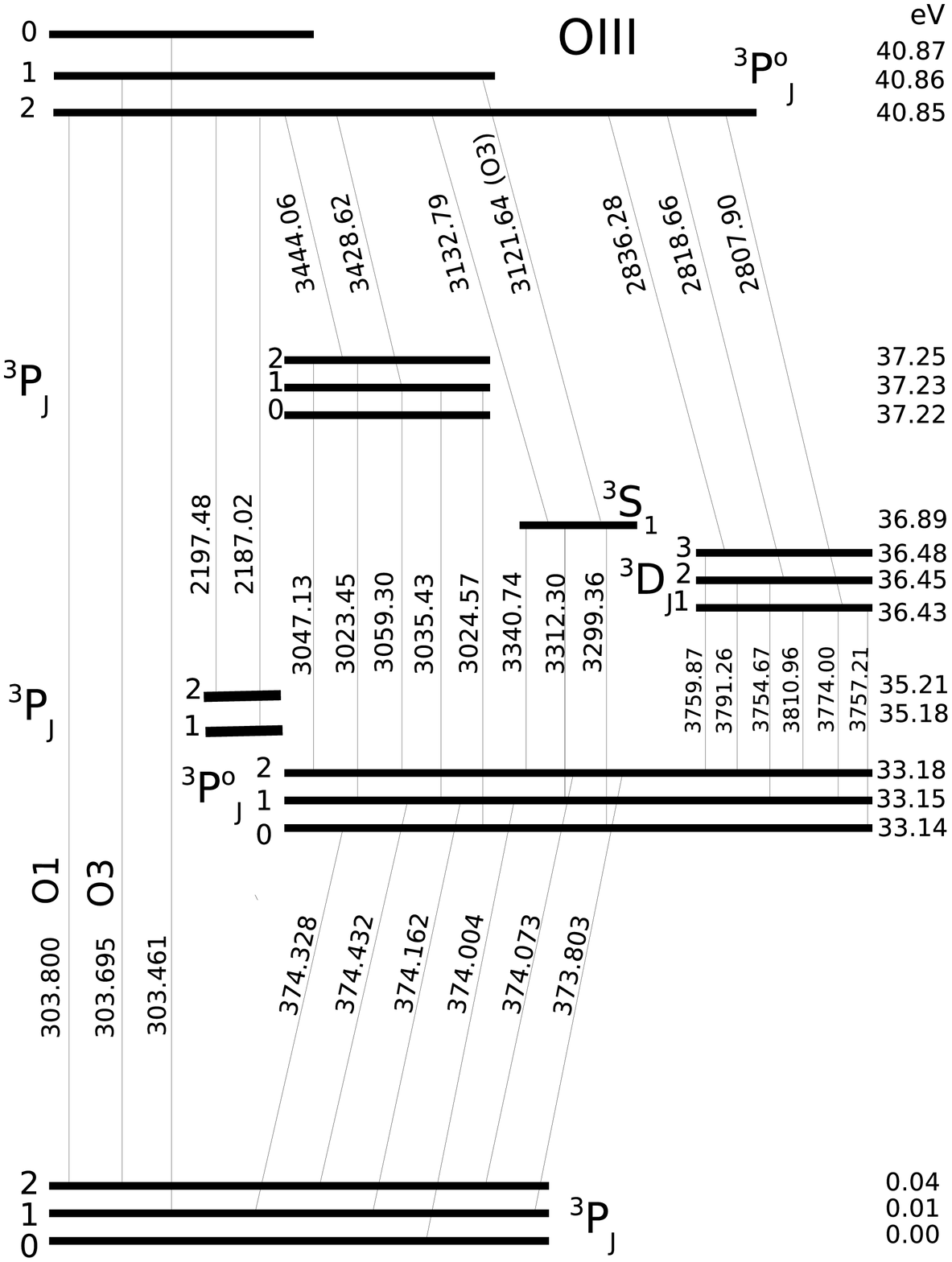}}
\caption{A  partial Grotrian diagram of $\ion{O}{iii}$ that includes 
the  most
relevant  O1 transitions and the $\lambda$ 3121.64 line belonging to the  
O3 process.  The level's configuration can be read from Table 3. The
y-axis
is  not 
scaled linearly. 
}
 \end{figure}

\begin{table}
\caption{ Configuration and  energy of the relevant $\ion{O}{iii}$
Bowen fluorescence levels.}
\begin{center}
\renewcommand{\tabcolsep}{0.8cm}
\begin{tabular}{lrr}
\hline
 Level   &       E(eV)  &  E(cm$^{-1})$     \\
2p3d $^3P_0^o$     & 40.87 &   329645.14    \\
2p3d $^3P_1^o$     & 40.86 &   329583.89    \\
2p3d $^3P_2^o$     & 40.85 &   329469.80    \\
2p3p $^3P_2$       & 37.25 &   300442.55    \\
2p3p $^3P_1$       & 37.23 &   300311.96     \\
2p3p $^3P_0$       & 37.22 &   300239.93     \\
2p3p $^3S_1$       & 36.89 &   297558.66    \\
2p3p $^3D_3$       & 36.48 &   294223.07     \\
2p3p $^3D_2$       & 36.45 &   294002.86     \\
2p3p $^3D_1$       & 36.43 &   293866.49     \\
2$p^4$ $^3P_1$     & 35.21 &   283977.4\phantom{0}    \\
2$p^4$ $^3P_2$     & 35.18 &   283759.7\phantom{0}    \\
2p3s $^3P^o_2$     & 33.18 &   267634.00     \\
2p3s $^3P^o_1$     & 33.15 &   267377.11      \\
2p3s $^3P^o_0$     & 33.14 &   267258.71     \\
2s2p$^3$ $^3S^o_1$ & 24.44 &   197087.7\phantom{0}  \\
2s2p$^3$ $^3P^o_0$ & 17.65 &   142393.5\phantom{0}     \\
2s2p$^3$ $^3D^o_1$ & 14.88 &   120058.2\phantom{0} \\
2s2p$^3$ $^3D^o_2$ & 14.88 &   120053.4\phantom{0} \\
2s2p$^3$ $^3D^o_3$ & 14.88 &   120025.2\phantom{0}  \\
2p$^2$ $^3P_2$     &  0.04 &      306.17 \\
2p$^2$ $^3P_1$     &  0.01 &      113.18  \\
2p$^2$ $^3P_0$     &  0.00 &        0.00  \\
\hline
\end{tabular}
\end{center}
\end{table}

In most planetary nebulae and symbiotic stars, the "efficiency" of the
Bowen mechanism, that is the fraction of $\ion{He}{ii}$ Ly-$\alpha$
photons that
is
converted into $\ion{O}{iii}$ Bowen lines (mostly a function of the
$\ion{He}{ii}$
Ly-$\alpha$
optical depth)  varies from object to object and  covers almost the
whole
range of possible values  (Liu and Danziger 1993, Pereira et al.,
1999).
Therefore, the Bowen mechanism can have a major influence on the fate of
the  $\ion{He}{ii}$ 
Ly-$\alpha$ photons and consequently also on the photoionization
equilibrium in the nebula. 
Thus, a clear understanding of the Bowen  mechanism is 
required for a correct  interpretation of the emergent UV and optical
spectrum.   

If the width of the $\ion{He}{ii}$ Ly-$\alpha$ line is large enough, an
additional
fluorescence process might  also be present i.e. the excitation of the
  $\ion{O}{iii}$   
(2p3d)$^3P_1$ level (v= -82 km/s from $\ion{He}{ii}$ $\lambda$ 303.782)
 due to the partial overlap of
the exciting  $\ion{He}{ii}$ Ly-$\alpha$  with the $\lambda$  303.695 
resonance line  of
$\ion{O}{iii}$  (O3 process). In addition, if  $\ion{He}{ii}$
Ly-$\alpha$ is even wider,
then
excitation of the $(2p3d) ^3P_0$ level by pumping in the
resonance line at $\lambda$  303.461  could  also take place. 
This is  a quite
uncommon process, since this line is at -250 km/s with respect  to 
$\ion{He}{ii}$ Ly-$\alpha$. Thus,
these two secondary Bowen fluorescence processes provide a convenient
probe which measures
the intensity of radiation in the far wings of the   
unobserved $\ion{He}{ii}$ Ly-$\alpha$ line
and  gives  valuable information on its width. It is worth recalling that
Bhatia et al.
(1982) in a study  of the solar $\ion{O}{iii}$ spectrum  have pointed
out the
fact that the $\ion{He}{ii}$ Ly-$\alpha$
line is considerably broader than the $\ion{O}{iii}$ lines. 

 It is important to  note that while the lines in the primary cascade
from
each  $(2p3d)^3P_{0,1,2}$ level are individual, (e.g. $\lambda$ 3132.79
from
$(2p3d)^3P_{2}$ (O1),  $\lambda$ 3121.64  from $(2p3d)^3P_{1}$ (O3), 
$\lambda$
3115.68 from  $(2p3d)^3P_{0}$)  most of the lines 
in the subsequent cascades are common to the three processes, although
the
dominant contribution comes from the O1 process.

We recall that  a competitive charge-exchange mechanism (CE) functions
for some
of the lines produced in the decay from the $^3D$ term  (Dalgarno and
Sternberg, 1982) ,  while
the  
$\lambda$  5592.25 line
(3s $^1P^o$-3p $^1P$, mult. 5)  comes from charge-exchange only (see
Liu and
Danziger 1993 and  Kastner and Bhatia 1996, hereinafter  KB96 for
details).

An accurate determination of the intensities of the pure Bowen lines
is
fundamental for determining the relative contribution  by the various
channels previously mentioned. KB96 have
pointed
out that a complete Bowen system has not yet been observed because of
inadequate  coverage in wavelength, lack of adequate  spectral
resolution, lack of
accurate ELI, and lack of overlap between the data of
various
observers. Also, the lack of sufficiently accurate intensities for the
weak CE line at $\lambda$ 5592.25, and for the  primary cascade line of 
the O3
process
at $\lambda$  3121.64 is particularly relevant, according to KB96.

\subsection{The Bowen lines intensities  in  RR Tel}

In our UVES data (spectral resolution $\sim$ 0.05 \AA~ FWHM) we have
detected and measured all of the $\ion{O}{iii}$ fluorescent lines (O1
and O3
processes) that are present in the range $\lambda$$\lambda$ 3047-3811,
with the sole
exception of the two lines at $\lambda$ 3759.87 and at $\lambda$ 3810.96
that
are blended with stronger lines of  [$\ion{Fe}{vii}$] $\lambda$ 3758.92
and
$\ion{O}{vi}$(1)
$\lambda$ 3811.35, respectively. The remaining lines are unblended
except for
the $\lambda$  3405.71 line which is partially affected by the presence
of
the $\lambda$ 3405.78
$\ion{O}{iv}$(3) line.

Figures  7 and  8 are self-explanatory and illustrate the exceptional
quality of the STIS-NUV-MAMA and UVES spectral data that have allowed the
detection and measurement of the pure lines produced in the various
Bowen excitation processes, including the very weak  $\lambda$
3115.68 line that is associated with  the excitation of the $(2p3d)
  ^3P_0$ level by the $\ion{He}{ii}$ $\lambda$  303.461  line.

\begin{figure}
\centering
\resizebox{\hsize}{!}{\includegraphics{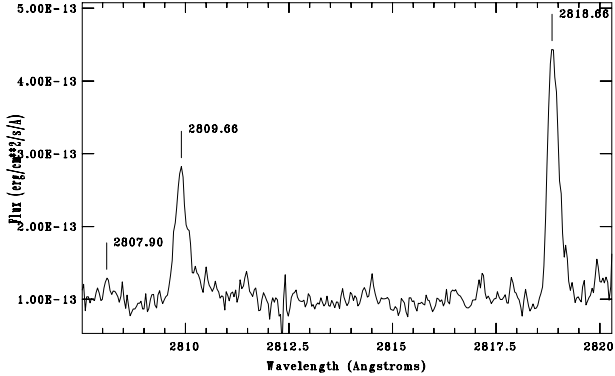}}
\caption{The   weak $\ion{O}{iii}$ Bowen lines  near
$\lambda$ 2800 in STIS }
 \end{figure}

\begin{figure}
\centering
\resizebox{\hsize}{!}{\includegraphics{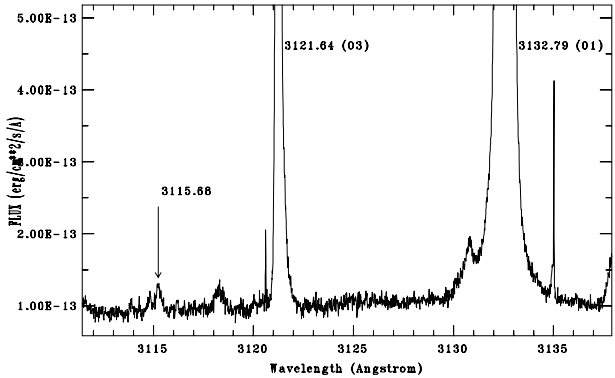}}
\caption{The UVES spectrum shows the strongest line of each of the three 
Bowen
excitation channels. From left to right: the very weak  $\lambda$
3115.68 line (from 2p3d $^3P_0$, at -250 km s$^{-1}$ from $\ion{He}{ii}$
Ly-$\alpha$),
the O3 $\lambda$ 3121.64 line  (from 2p3d $^3P_1$, at -88.2 km s$^{-1}$
from $\ion{He}{ii}$ Ly-$\alpha$), and the strong O1 $\lambda$ 3132.79
line (from
2p3d $^3P_2$, at +16.4 km s$^{-1}$ from $\ion{He}{ii}$ Ly-$\alpha$).
}
\end{figure}

The STIS data are complementary to the UVES data shortward of $\lambda$
3047 (see also Fig. 9) and include all of the Bowen lines present in
this range,
although 
the  two lines at $\lambda$ 2807.90   and  $\lambda$  2798.90   are very
weak.  In the
spectral range that is common with UVES (longward of $\lambda$ 3046)
the
STIS data include  most  of the Bowen lines detected by UVES except 
the weakest ones that  have not been detected because of lower
spectral resolution and S/N in the STIS optical  data  compared with 
that
of
UVES.

The merging of the STIS and UVES data (as mentioned in Sect. 2)
has thus provided a high quality  coverage of ELI for
almost all of
the $\ion{O}{iii}$ Bowen lines in the range $\lambda$ $\lambda$
2798-3810.

Table 4 gives the observed STIS and UVES  ELI (10$^{-13}$ erg cm$^{-2}$
s$^{-1}$) and their  ratios with respect to the
reference line $\lambda$ 3444.06. The wavelengths and the radiative
rates A$_{ij}$  are 
 from KB96, that adopted the IC rates of Froese Fisher (1994). 
It should be noted that these values differ by about 10 $\%$, 
with nearly constant ratio, from those listed in the NIST 
 Web site, whose accuracy is in the  range   C ($\leq$ 25 $\%$) -
 C$^+$ ($\leq$ 18 $\%$). Therefore, the level branching 
ratios are nearly the same, and the detailed discussion of the decays
(see also 
Section 4.6) is not  affected. 
We note,
incidentally,  that the  $\lambda$  2808.77 line in KB96 should
be correctly reported as  $\lambda$  2807.90.
In this list we have also included
for
comparison the CE process line at $\lambda$ 5592.25, the hydrogen
H$_{\beta}$
line $\lambda$ 4861, and the $\ion{He}{ii}$ Ba-$\alpha$ $\lambda$ 1640,
Pa-${\beta}$
 $\lambda$  3203 and
Pa-$\alpha$ $\lambda$  4685.71 lines. The FWHM values (deconvolved from
the
instrumental FWHM) are from STIS in the
UV
region (shortward of $\lambda$ 3044 ) and from UVES in the optical
region, with
the only exception of the FWHM of $\ion{He}{ii}$ $\lambda$ 4686 that is
from STIS.
The last column of Table 4 gives the number of photons in the line,
obtained from the observed ELI after proper conversion.

\begin{table*}
\caption{ Bowen Fluorescence Lines. The nine colums give: the
laboratory wavelengths (\AA, in
air), the  transition, the A$_{ij}$
(s$^{-1}$) 
(from KB96, except for the
first two lines taken from "The 
Atomic Lines List 2.05" (http://www.pa.uky.edu/$\sim$peter/newpage);
1.48e+8 represents
1.48$\times$10$^{+8}$), the  line  FWHM (km s$^{-1}$), the ELI
($10^{-13}$erg
cm$^{-2}$s$^{-1}$) from STIS and UVES data,  
 the  relative ELI STIS$_r$  and UVES$_r$
(I$_{3444.06}$=100),  and the photon number
N$_{\nu}$  (in
10$^{-3}$ photons cm$^{-2}$s$^{-1}$) from the STIS scale. }
\begin{center}
\renewcommand{\tabcolsep}{0.3cm}
\begin{tabular}{lcccccccc}
\hline
$\lambda_{air}$ &  Transition  & A$_{ij}$ & FWHM   & STIS & UVES  &
STIS$_r$
&
UVES$_r$ &
N$_{\nu}$\\
 \hline
2187.02 & 3d $^3P_2^o$-2$p^4$ $^3P_2$  &7.19+5        & (41.3)  &0.40     
&   -  
&   1.24  
&  
-  
&
4.41 \\
2197.48 &  3d $^3P_2^o$-2$p^4$ $^3P_1$  &2.36+5       & (40.5)  & 0.21    
&   -  
&    0.65 
&   
-  
&
2.32  \\
2798.90 & 3d $^3P_1^o$-3p $^3D_1$ &3.14+6        &  -       &    (0.05)   
&   -  
& 
(0.15)& -   
&
(0.70)  \\                                                           
2807.90 &  3d $^3P_2^o$-3p $^3D_1$   &1.79+5      &  -       &(0.03)       
&  
-     &  
(0.1)  &- 
&
(0.42) \\                                                                   
2809.66 &  3d $^3P_1^o$-3p $^3D_2$ & 1.41+7       & (39.3)  &0.65      &  
-    
&   2.02  
&    - 
&
9.21 \\                                                                
2818.66 &  3d $^3P_2^o$-3p $^3D_2$   &  1.57+6       & 33.0    &1.12     
&   -    
&  
3.48   &    - 
&
14.47\\                                                                
2836.28 &   3d $^3P_2^o$-3p $^3D_3$   & 1.64+7       & 35.7    &12.20    
&    -   
&  
37.89  &    - 
&
174.19\\                                                                   
3023.45 &  3p $^3P_2$-3s $^3P^o_1$  &5.17+7      & 37.1    &2.37      &  
-    
&    7.36 
&   
-  & 
36.07\\                                                                                                                               
3024.57 &  3p $^3P_1$-3s $^3P^o_0$  & 6.73+7       & (43.5)  &1.71      & 
-     
&   
5.31 
&    - 
&
26.04 \\                                                                                                                                              
3035.43 &    23p $^3P_1$-3s $^3P^o_1$  & 5.06+7       & 34.1    &0.87     
&   -    
&   
2.70 
&    - 
&
13.29 \\                                                                                                                                                            
3043.02 &    3p $^3P_0$-3s $^3P^o_1$ & 2.14+8      & 28.6    &0.37      &  
-    
&   
1.15  &     - 
&
5.66 \\                                                                                   
3047.13 &  3p $^3P_2$-3s $^3P^o_2$  & 1.63+8      & 34.8    &15.30     & 
13.91 
&   47.52 
&  49.86 
&234.70  \\                                                           
3059.30 &    3p $^3P_1$-3s $^3P^o_2$  & 9.46+7      & 34.5    &1.74     
&  1.44  
&   
5.40 
&   5.16 
&
26.80 \\                                                             
3115.68 &   3d $^3P_0^o$-3p $^3S_1$ &1.41+8       & 36.4     & -      & 
(0.11)
&      - 
&    (0.39) 
&
(2.04) \\                                                                   
3121.64 &    3d $^3P_1^o$-3p $^3S_1$  &  1.44+8     & 27.9   &3.67      &  
3.25 
&  
11.40  &   11.63 
& 
57.65\\                                                                   
3132.79 &   3d $^3P_2^o$-3p $^3S_1$ & 1.52+8     & 34.6   &102.4    &  
86.9 & 
318.0  &  
311.5 
&
1615.1 \\                                                 
3299.36 &  3p $^3S_1$-3s $^3P^o_0$  &1.79+7    & 34.2    & 4.32      &   
3.66 &   
13.42
&  
13.12 
&
71.76\\                                                                       
3312.30 &   3p $^3S_1$-3s $^3P^o_1$  & 4.97+7      & 34.1   &11.40     &  
9.50 
&    35.40
&   34.05 
&
190.01\\                                                              
3340.74 &  3p $^3S_1$-3s $^3P^o_2$  &6.84+7     & 34.1    &15.10     &  
12.85 &   
46.89
&   46.09 
&
253.90\\                                                                     
3405.71 & 3d $^3P_1^o$-3p $^3P_0$ &1.78+7      & 28.9    &-         &  
0.40  &     
-  
&   1.43 
&
8.23 \\                                                                    
3408.12 &   3d $^3P_0^o$-3p $^3P_1$   &7.27+7    & 25.3    &-         & 
(0.04)
&       - 
&   (0.14) 
&
(0.85) \\                                                          
3415.26 & 3d $^3P_1^o$-3p $^3P_1$    &   2.31+7       & 27.8   &-        
&  
0.49  &     
-   &   1.76 
&
10.31 \\                                                           
3428.62 &   3d $^3P_2^o$ -3p $^3P_1$   &7.98+6     & 38.2    &4.74      &   
3.78 &   
14.72 &   13.55 
&
81.79 \\                                                        
3430.57 &     3d $^3P_1^o$-3p $^3P_2$ & 2.78+7      & 26.7    &(0.60)   
&  
0.55  &  
(1.86) &  1.97 
& 
10.38\\                                                                
3444.06 & 3d $^3P_2^o$-3p $^3P_2$  &   5.21+7   & 34.9    &32.20      &  
27.90
&   100.0
&   100.0 
&
558.25 \\                                                                    
3754.67 & 3p $^3D_2$-3s $^3P^o_1$   &   8.67+7      & 31.8   &(0.77)    &  
1.05 
&   
(2.39)  &   3.76 
&
14.55 \\                                                                   
3757.21 &  3p $^3D_1$-3s $^3P^o_0$ & 6.42+7     & 33.5    &(0.25)    & 
0.24   &   
(0.78) 
&   
0.86 
&
4.73 \\                                                                  
3759.87 &    3p $^3D_3$-3s $^3P^o_2$& 1.13+8       &  -      & bl      &   
bl  
&    bl   
&    bl 
& -
\\                                                                          
3774.00 &    3p $^3D_1$-3s $^3P^o_1$ &4.55+7      & 30.7    &(0.28)    &  
0.22 
&  (0.87) 
&  0.79 
&
5.23 \\                                                                                
3791.26 &    3p $^3D_2$-3s $^3P^o_2$  &2.61+7       & 35.4    &-        &  
0.30 
&      -  
&   1.08 
&
6.87  \\                                                                          
3810.96 &    3p $^3D_1$-3s $^3P^o_2$  & 2.79+6       &  -      & -       
&    
-   &      
-  &    - 
& -
\\                                                                             
5592.25   & 3s $^1P^o$-3p $^1P$ &-        &  37.3   & -        &    0.12
&     
-   &  
0.44 
& 
4.05\\                                                                         
$\ion{He}{ii}$-3203  & 3-5 &-     &  54.2   &63.5       &   54.65 & 
197.2 
&   195.9 
&
1023.9 \\                                                                             
$\ion{He}{ii}$-4686  & 3-4 &-      &  56.2    &147.5      &    -    &  
458.0  &    - 
&
3477.9 \\                                                                          
$\ion{He}{ii}$-1640   & 2-3   &-       &  63.2   &831.0    &   -     & 
2580.7  &    - 
&
6860.4 \\                                                                            
HI-4861 & 2-4  & -          & 60.0 &162.0 &   -     &         503.1 &      
- 
&
3964.7 \\ 

\hline
\end{tabular}
\end{center}
\end{table*}

As mentioned in section 2.2,  the  intensity ratios  for the lines that
are common to the two sets of
measurements are in  good agreement with mean value = 1.16 and variation
coefficient = 2.9 $\%$.
The consistency between the two sets of data lend support to the
re-calibration method adopted for  UVES  and to the overall correctness
of the method.

It should be pointed out that in an earlier short study based only on
UVES data (Selvelli \& Bonifacio 2001) the values for the relative
emission intensities were slightly different since the UVES calibration
was based on standard ground-based reduction method, and the correction
for the color excess was assumed to be (E$_{(B-V)}$=0.10. However, only
for
the $\lambda$ 5592.25 line is the difference with the previous
measurements significant.

The I$_{3133}$/I$_{3444}$ ratio in STIS data is = 3.18, and,
consistently = 3.11 in
UVES data. The same ratio was reported as = 3.39 by Pereira et al
(1999), in a study of Bowen fluorescence lines
for a group of symbiotic stars, including RR Tel  using ESO data with
spectral resolution of $\sim 2.5$ \AA~ (FWHM). As mentioned in section
3.4 they have adopted E$_{(B-V)}$=0.04.

In conclusion, we have obtained high quality data for about thirty
$\ion{O}{iii}$
lines of the  Bowen  fluorescence process, including six pure O1
lines  ($\lambda$$\lambda$ 2807.90, 2818.66, 2836.28, 3132.79, 3428.62
and  3759.87),   
seven
pure O3 lines ($\lambda$$\lambda$    2798.90, 2809.66, 3043.02, 3121.64,
3405.71, 3415.26, and 3430.57) and the two pure very weak lines at
$\lambda$ 3115.68 
 and $\lambda$ 3408.12  that come from the pumping of level 2p3d
$^3P_0^o$  
(40.87  eV) which  requires a  very wide $\ion{He}{ii}$ Ly-$\alpha$ 
line  
(this process is called "other" in KB96).  It is worth recalling
(see
also KB96) that these two lines
were reported (but not measured) in the comprehensive paper on RR Tel
by Thackeray (1977). The $\lambda$ 3115.68  line was correctly
identified
as
$\ion{O}{iii}$, while the $\lambda$ 3408.12 line was tentatively
identified as a
CrII
line.

\begin{figure}
\centering
\resizebox{\hsize}{!}{\includegraphics{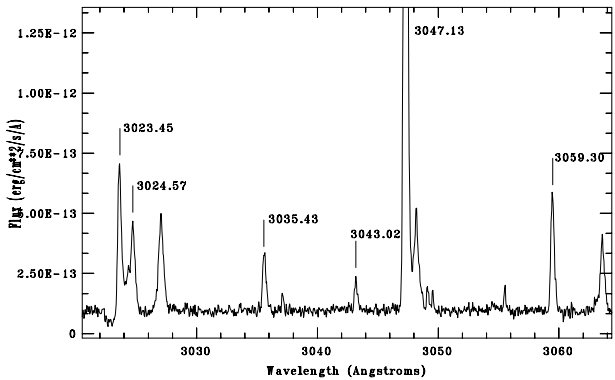}}
\caption{The   weak $\ion{O}{iii}$ Bowen lines  in the   STIS region
near
$\lambda$ 3030.}
 \end{figure}

\subsection{The Bowen lines width and profile}

The  data for the FWHM  for the $\ion{O}{iii}$ Bowen lines reported in
Table 4 
 (obtained from the high resolution spectra of STIS and UVES) give  an
average FWHM value of 34.30 $\pm$ 1.90 km s$^{-1}$ for
the O1 + mixed (O1+O3) lines, while the average FWHM for the five best
observed pure
O3 lines is = 27.98 $\pm$ 0.85 km s$^{-1}$. The difference is 
significant
with
a ratio
FWHM(O1)/FWHM(O3)  $\sim$ 1.22.

It is also  remarkable that  in UVES spectra all the best observed
$\ion{O}{iii}$
Bowen lines
(i.e. $\lambda$ 3133,  $\lambda$ 3121, $\lambda$  3299,  $\lambda$ 3312,
$\lambda$  3340,
$\lambda$ 3444, etc.) show a definite asymmetry (excess) in the red
wing (see Fig. 10). 
The red "excess" (after
subtraction of the main gaussian component of about 34.3 km s$^{-1}$
FWHM)
can be modeled by a
gaussian profile centered at about +38 km s$^{-1}$ from the main
component
and with FWHM of about 22 km s$^{-1}$. There is however still an
"excess"
left
and
the velocity profile in the "red" wing can reach a maximum velocity of
$\sim$ 100 km s$^{-1}$. Alternatively, the entire profile can be also
(and
better) fitted with a gaussian main component with FWHM 31 km s$^{-1}$m,
while the red "excess" is represented by a Voigt profile with FWHM 25 km
s$^{-1}$  in the
Gaussian core and 17 km s$^{-1}$ in the Lorentzian wing.

\begin{figure}
\resizebox{\hsize}{!}{\includegraphics{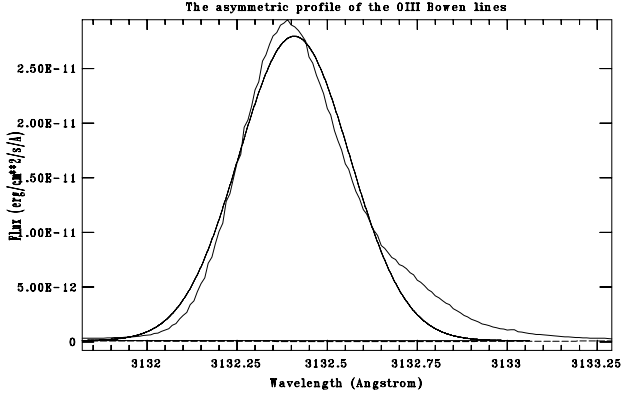}}
\caption{ The  asymmetric profile of the $\ion{O}{iii}$ $\lambda$ 3133
line and a gaussian fit with FWHM=34.5 km s$^{-1}$.   The
same kind
of
 profile is present also in other lines, see sect. 4.3. }
 \end{figure}

A similar but less enhanced behavior is shown by the $\ion{He}{i}$ lines
at
$\lambda$
5875, and $\lambda$ 6678, while the two [$\ion{O}{i}$] lines at
$\lambda$ 6300 and
$\lambda$ 6363 show two
almost resolved components with separation of 27 km s$^{-1}$ and FWHM of
13
km s$^{-1}$ and 28 km s$^{-1}$ respectively.

Surprisingly, the nebular [$\ion{O}{iii}$] lines at $\lambda$ 4958 and
$\lambda$
5007 show instead an
excess in the blue wing with a maximum velocity of about -150 km
s$^{-1}$,
while other strong emission lines observed with UVES e.g.: $\ion{He}
{ii}$
$\lambda$ 3203, [$\ion{Ne}{v}$]
$\lambda$ 3426, [$\ion{Fe}{vii}$] $\lambda$ 3586 , H-$\beta$ $\lambda$
 4861,  [$\ion{Fe}{vii}$]
$\lambda$ 4893, [$\ion{Ca}{vii}$] $\lambda$ 4939, [$\ion{Fe}{vii}$]
$\lambda$ 5176, $\ion{He}{ii}$ $\lambda$ 5411, [$\ion{Ca}{vii}$]
$\lambda$ 5618,  [$\ion{Fe}{vii}$] $\lambda$ 5721, [$\ion{Fe}{vii}$]
$\lambda$ 6086, [$\ion{Ar}{v}$] $\lambda$ 6434, and [$\ion{Ar}{v}$]
$\lambda$ 6455 show  an
almost pure Gaussian profile.

We recall that Crawford (1999) reported the presence of a two
component structure in the [$\ion{O}{iii}$] $\lambda$ 4363 line and of a
multi-component
structure in the [$\ion{O}{iii}$] $\lambda$ 4959 and  $\lambda$ 5007
lines.
Schild and Schmid (1996) also observed this behavior
and attributed it to the presence of two nebular components of different
densities.

In the far UV and near UV   echelle spectra  of STIS (taken 1 year after
the
UVES spectra) whose resolution is comparable to that of UVES, a red wing
excess
similar to that present in  $\lambda$ 3132.79 is present in all of the
common
intercombination lines , e.g. $\ion{O}{v}$] $\lambda$ 1218,
 $\ion{O}{iv}$] $\lambda$ 1404,
$\ion{N}{iv}$] $\lambda$ 1486, $\ion{O}{iii}$] $\lambda$ 1660 and
$\lambda$ 1666,  $\ion{N}{iii}$] $\lambda$ 1750,
$\ion{Si}{iii}$]
$\lambda$
1892 and $\ion{C}{iii}$] $\lambda$ 1906. It is instead
absent in the lines of permitted transitions, i.e. in the three
resonance doublets of $\ion{N}{v}$ ($\lambda$  1238, $\lambda$ 1242),
SiIV
($\lambda$1293, $\lambda$ 1403), and $\ion{C}{iv}$ ($\lambda$ 1548,
$\lambda$
1550)
and in
the $\ion{He}{ii}$ Ba-$\alpha$ recombination line $\lambda$ 1640,  all
of which
display
a
nearly pure Gaussian profile

The asymmetry is not evident in the STIS optical [$\ion{O}{iii}$]
 lines
possibly
because of the lower spectral resolution of STIS in the optical range.
In FEROS spectra, in the range outside that covered by UVES, there
is
evidence of excess in the  [$\ion{O}{iii}$] $\lambda$ 4363 nebular line
(as
mentioned
by
Crawford, 1999), and in the $\ion{O}{i}$  $\lambda$ 8446  line produced
by the
Ly-$\beta$ fluorescence. No excess is evident in H$_{\gamma}$ $\lambda$
4340, and in $\ion{He}{ii}$  $\lambda$ 4686.

\subsection {The relative Bowen line intensities and the  efficiency of
the
O1 and O3 processes}

The high quality of the data has allowed a detailed comparison  of  the
observed  ratios for pairs of lines from a common upper level to the
corresponding branching 
ratios obtained from  the  transition rates given by   Bhatia and Kastner
(1993)
(BK),
Froese Fischer (1984) (FF), and Saraph and Seaton (1980) (SS).

Following Pereira et al (1999) we
present in Table 5 a comparison between the
observed and the predicted  ratios for some selected lines. It is
clear from Table 5  that the best agreement in the
overall ratios is
with the Froese Fisher (1994) model. This strongly supports the  
intermediate coupling radiative rates calculated by FF.

\begin{table}
\caption{A comparison between observed and
predicted intensity ratios for pairs of  Bowen lines originating from
the
same upper level. The  theoretical intensity ratios  have been derived
from  the transition rates of Bhatia and Kastner (1993) (BK), 
Froese-Fisher (1994) (FF), and Saraph and Seaton (1980) (SS). }
\begin{center}
\renewcommand{\tabcolsep}{0.2cm}
\begin{tabular}{lrrrrr}
\hline
line ratios &  STIS  &  UVES  & BK   &  FF   &  SS   \\
2819/3444   &  0.035 &   -   & 0.033 & 0.037 &   -   \\
2836/3444   &  0.38  &   -   & 0.43  & 0.38  &   -   \\
3133/3444   &  3.18  & 3.12  & 4.06  & 3.21  &  3.61 \\
3429/3444   &  0.15  & 0.14  & 0.18  & 0.15  &  0.34 \\
3406/3415   &   -    & 0.81  & 0.75  & 0.78  &   -  \\
3415/3122   &   -    & 0.15  & 0.09  & 0.15  &   -   \\
3299/3341   &  0.29  & 0.28  & 0.23  & 0.26  &  0.20  \\
3299/3312   &  0.38  & 0.38  & 0.35  & 0.36  &  0.34  \\
3791/3755   &   -    & 0.29  & 0.31  & 0.30  &  0.33  \\
\hline
\end{tabular}
\end{center}
\end{table}

The efficiency of the Bowen fluorescence mechanism is quantified by the
fraction of the created $\ion{He}{ii}$ Ly-$\alpha$ photons ($\lambda$
303.782)
that
are converted  into Bowen line photons. If the atomic
transition probabilities are known, the intensity of any $\ion{He}{ii}$
line can
be related to the intensity of the $\ion{He}{ii}$ Ly-$\alpha$ line (and
to the
number of $\ion{He}{ii}$ Ly alpha photons) and the intensity of any
Bowen line
can be related to the total intensity of the Bowen lines (and total
number of Bowen photons). In this case, the efficiency can be easily
calculated if the (relative) intensities of a single Bowen line
(usually $\lambda$ 3133 or $\lambda$ 3444) and a single $\ion{He}{ii}$
line
(usually
$\lambda$ 4686, or $\lambda$ 3203)  are available. The efficiency factor
R can be defined as:
\begin{displaymath}R=\frac{P \cdot \alpha_{eff}(4686) \cdot \lambda
\cdot
I_{\lambda}}{P_{\lambda} \cdot \alpha_{eff}(304) \cdot 4686 \cdot
I_{4686}}\end{displaymath}
where $\alpha_{eff}(304)$ and $\alpha_{eff}(4686)$ are the effective
recombination coefficients  for recaptures that result in the production
of $\lambda$ 304 and $\lambda$ 4686 respectively, P is the cascade
probability =
0.0187, and P$_{\lambda}$ is the probability for the emission of a
particular Bowen line following the excitation of 2p3d $^3P_2^o$ (O1) 
(Aller, 1984). The ratio of the two $\alpha_{eff}$ is
0.328
for T$_e$=10,000 K. 

The application  to specific lines provides simple  relations that can
be
directly used, i.e.  R$\sim$1.0 $\cdot$ I$_{3444}$/I$_{4686}$
(Kaler 1967,  Harrington 1972), R=0.12 $\cdot$ I$_{3133}$/I$_{3203}$
(Schachter et al. 1990), 
R=0.32 $\cdot$ I$_{3133}$/I$_{4686}$
(Saraph and Seaton 1980) R=0.43 $\cdot$ I$_{3444}$/I$_{3203}$
(Wallerstein, 1991). 
In the specific case of RR Tel these relations give R = 0.22, 0.21,
0.22,
and 0.22, respectively, in encouraging agreement.

Similar relations can be obtained for the efficiency of the O3 channel.
Shachter et al. (1991) give

\begin{displaymath}\frac{R_{O3}}{R_{O1}}=\frac{I_{3122} \cdot A_{3133}
\cdot 3122}{I_{3133} \cdot A_{3122} \cdot 3133}\end{displaymath}
and   for the specific case of RR Tel one obtains  that the relative 
R$_{O3}$/R$_{O1}$ efficiency is  $\sim$ 0.038  (a value that is much
lower than that, $\sim$
0.3, found  by Shachter et al., 1991 for  AM Her).  Therefore,  the 
efficiency  of the O3 channel in RR Tel is close to 0.7 $\%$.

In RR Tel we have obtained the intensities of 
all the Bowen lines  belonging  to the primary cascade. This fortunate 
situation  allows a  direct counting of  the number of Bowen
photons  observed in the 6 lines of the primary O1 channel (the
$\lambda$
2808.77 line 
is not detected)  and therefore  a more  direct estimate of the
efficiency.  
The total intensity in the 6 lines of the primary O1 decay from level  
2p3d $^3P_2^o$ (O1) is of 152.7 10$^{-13}$ erg cm$^{-2}$s$^{-1}$
(from STIS  data).
This value, after proper conversion from energy to  number of photons,
corresponds to a total number of photons (decays) in the O1 primary
cascade of 
approximately  2.45 photons cm$^{-2}$ s$^{-1}$  (see also Table 4).

The $\ion{He}{ii}$ Ly-$\alpha$ intensity can be estimated from the
observed
$\lambda$ 4686
line intensity ($\sim$ 147.5 10$^{-13}$ erg cm$^{-2}$ s$^{-1}$) assuming 
a ratio
I$_{304}$/I$_{4686}$ $\sim$ 65 (Storey and Hummer, 1995). The
corresponding 
number of $\ion{He}{ii}$ Ly-$\alpha$ photons is about 15.36 cm$^{-2}$
s$^{-1}$.
Thus, the efficiency for the O1 channel R(O1) is close to 16.0 $\%$, 
with some uncertainty associated with the actual value of the  $\ion{He}
{ii}$ Ly-$\alpha$ intensity.

We recall that the excitation of the 2p3d $^3P_1$ level of
$\ion{O}{iii}$ (the O3 channel) requires a velocity
shift of
-88.2 km/s from the rest wavelength of the $\ion{He}{ii}$ Ly-$\alpha$
line,
while
the excitation of the 2p3d $^3P_0$ level  (that produces the two
very weak but observed lines at $\lambda$
3115 and
$\lambda$ 3408) 
requires a corresponding velocity shift of -250 km/s. This clearly
indicates that the  $\lambda$ 303.782  $\ion{He}{ii}$ Ly-$\alpha$  line   
must
be broad  enough to excite these levels of $\ion{O}{iii}$. 
 
As reported in Sect. 3.1,  the average FWHM for 20 unblended
lines of the Fowler series is = 53.5 $\pm$ 3.5 km s$^{-1}$ .
while  the $\ion{He}{ii}$  Ba-$\alpha$ line $\lambda$  1640  has FWHM =
63.2 km
s$^{-1}$. These
values
are
about ten times larger than the thermal velocities that are on the
order of about 6.4 km s$^{-1}$.

It is clear from these values that the $\ion{He}{ii}$ Ly-$\alpha$ line
 is
much broader than the $\ion{He}{ii}$ recombination lines, probably
 a consequence  of 
multiple resonant scatterings.

\subsection{The past efficiency of the $\ion{O}{iii}$ Bowen fluorescence
process }

In most IUE LW high resolution spectra of RR it is possible to obtain
good measurements for the $\ion{He}{ii}$ Fowler lines at $\lambda$ 2733
and $\lambda$ 3203
 and for the $\ion{O}{iii}$ lines at $\lambda$ 2836, $\lambda$  3047 and
$\lambda$ 3133 (O1 process) and the $\ion{O}{iii}$ line 
at $\lambda$  3122 (O3 process). Since the IUE spectra cover about 17
years
in
the
life of RR Tel one can thus  follow  the changes with time in the
absolute and
relative intensities of the $\ion{He}{ii}$ and $\ion{O}{iii}$ lines and
estimate the
corresponding changes in the Bowen efficiency. Table 6 gives the time
variation from 1978 to 1995 for these lines, as obtained from
measurements on more than 30 IUE spectra,  together with  the data  for
Oct. 10,2000 from STIS (last line).

\begin{table}
\caption{The  intensity  (in 10$^{-13}$erg cm$^{-2}$ s$^{-1}$)
of the $\ion{He}{ii}$  and $\ion{O}{iii}$ emission lines from IUE high
resolution 
spectra and STIS.   MJD  is the Modified Julian Date =
JD - 2,400,000.5.}
\begin{center}
\renewcommand{\tabcolsep}{0.1cm}
\begin{tabular}{lrrrrrrr}
\hline
IUE image  & MJD  &  $\lambda$2733  & $\lambda$ 2836 & $\lambda$3047  
&$\lambda$3122 &$\lambda$3133 &$\lambda$3203 \\                                                
LWR02021&43728  &     695 &    684 & 633   &330   &4610  &  1520\\
LWR02995&43833    &   700   &  520  & 742 & 273  & 4345 &   1398\\
LWR03888&43932    &   762  &  594 &  627 &  255   &3965  &  1133\\
LWR07536&44329  &     593 &   398   &619  & 255   &3715 &  1091\\
LWR07663&44363  &    667  &  526   & 603 &  254   &4239  & 1157\\
LWR08234&44433  &    698  &  518&    639 &   282 &  4203   &1213\\
LWR08272&44437  &    681  &  562&    631 &   247   &3987 &  1370\\
LWR09238&44548   &   650  &  454  &  635  &  210  & 3802  & 1029\\
LWR10364&44710    &  570 &   508&    479   & 210  &  3553  & 1032\\
LWR11293&44827&     669   &  480&    605  &  162  &  4268    &1187\\
LWR11296&44827    &  687 &   436 &   618 &   238   &  4348   &1141\\
LWR11744&44887    &  672   & 495 &   629 &   196  &  3147   & 1140\\
LWR14466&45265    &  632  &  486  &  610  & 181    &3792  &  1432\\
LWP01664&45229    & 668  &  473  &  537  &   172  &  3963&   1167\\
LWR16181&45503   &  564 &   411   & 471  &  181   &  3464  &  1003 \\
LWP03001&45780  &   619 &   475&    460   & 173 &   3753   & 1051\\
LWP08178&46561 &   629   &  419&    429   &  170 &   3313  &  1088\\
LWP08180&46562 &   578 &    459  &  460   & 155   & 2876   & 1074\\
LWP08728&46635 &   602 &   378  &  425   & 128  &  2940     &1136\\
LWP08729&46636 &   575  &  379 &  408    & 141&    3260    &  1156\\
LWP10919&46951  &  617 &   391&    409&    131  & 2990   &  1118\\
LPW13770&47374 &   566  &  329 &  326  &    95   &  2197   & 1042\\
LWP20538&48414  &  541  & 343    &293   &  86  &    2110  &   1107\\
LWP23569&48826 &   483   &  276  &  301  & 88    &  2022 &   1100\\
LWP24278&48932 &    567  &  268   & 319   & 85  &  1970  &    835\\
LWP25953&49188 &    606  &   284  &  354  & 81   & 2078    & 1055\\
LWP25954&49188 &    598 &   288   & 335  & 106  &  2225   & 916\\
LWP28078&49479 &   583 &    298 &   334  &  84   & 2238   &  1030\\
LWP29189&49611 &   586  &  273   & 297   &  96  &  2030  &   939\\
LWP30848&49872 &    561&    320  &  290   & 90  &  1964   &  1030\\
LWP30849&49872 &   516 &   229   & 286   & 72  &   1884   &  874\\
LWP31348&49953 &   548 &   278 &    241 &   59&    1767    & 979\\
STIS&51827&  323 &   122    & 155 &   37 &   1024    & 635\\

\hline
\end{tabular}
\end{center}
\end{table}

As already mentioned in Sect. 2.3, the   data  in    Table 6 all come 
from a choice of  good quality spectra. 
The data clearly show that starting from the first spectra secured in
August 1978 until the last spectra of August 1995 there has been a
steady decrease with time in all ELI.

The data have been fitted with both linear and power-law regression.
The  behavior with time  of the  two
$\ion{He}{ii}$
lines,   
that of  the three $\ion{O}{iii}$ O1 lines   and that   of  the $\ion{O}
{iii}$  O3 line   is
described by  different  power-law   indices ($\sim$  -0.33,
$\sim$ -0.97,    and   $\sim$ -1.48,  respectively)  but  within  each
line  group  the  indices are similar.
Figure 11 is a log-log plot of the data with their power-law fittings
(straight lines) for the 6 spectral lines listed in Table 6.

\begin{figure}
\resizebox{\hsize}{!}{\includegraphics{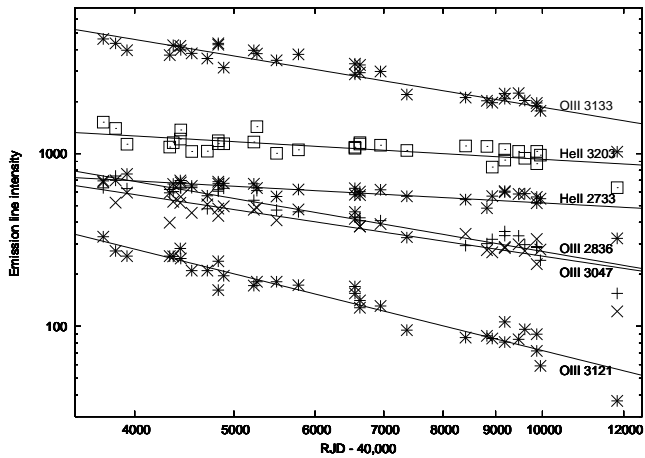}}
\caption{A log-log plot of  the  decrease with time  of the ELI  of the
$\ion{He}{ii}$
 and $\ion{O}{iii}$ lines from  1978 to 1995 in IUE spectra. The last
data  are
 from  STIS  (Oct.10, 2000).  The ELI are in units of 
10$^{-13}$ erg cm$^{-2}$ s$^{-1}$. The  
power-law 
fits to the data  (straight lines) have  indices $\sim$  -0.33,
$\sim$ -0.97,  and   $\sim$ -1.48,  for the two $\ion{He}{ii}$
lines  ($\lambda$ 2733 and $\lambda$ 3203),   the three $\ion{O}{iii}$
O1 lines  
($\lambda$ 2836,
$\lambda$
3047 and $\lambda$ 3132), and the  $\ion{O}{iii}$  O3 line ($\lambda$
 3122), 
respectively.}
 \end{figure}

On average, with the power-law fit, the intensity of the $\ion{He}{ii}$
lines (
$\lambda$ 2733 and $\lambda$ 3203) has decreased by a factor about 1.5
from 1978
to 2000. Instead, the intensity of the O1 lines ($\lambda$ 2836,
$\lambda$
3047 and $\lambda$ 3132)  has decreased on the average by a factor
larger
than 3, and the O3
line ($\lambda$  3122) has decreased by a factor larger than 5.

Therefore, the efficiency in the O1 channel (as determined by the
I$_{3133}$/I$_{3203}$ ratio) from the power-law  fitting was about 2.1
times higher
in
the early IUE years (1978-1980) and has decreased from a value near
0.5 in 1978-1980 to a value close to 0.2 in year 2000 ( STIS ).
Similarly, the data indicate that the O3 efficiency has decreased more
rapidly with time than the O1 efficiency and that in 1978-1980  the
relative O3  over O1 efficiency was about two times higher than in
2000.

 The O3  line is at -88.2 km/s from the
 $\ion{He}{ii}$  Ly-${\alpha}$ line.
The higher relative efficiency O3/O1 in the past can be explained by
a larger   width  in the $\ion{He}{ii}$ Ly-$\alpha$ line,  resulting
from
either 
a larger  turbulence or larger  optical depth effects 
 in epochs closer to the outburst time. 
We attribute the decrease  with time in the $\ion{He}{ii}$ ELI
to the general decrease in
the luminosity of the central source (Murset and Nussbaumer, 1993).
The corresponding stronger decline in the $\ion{O}{iii}$ Bowen lines
indicates a 
gradual decrease in the Bowen efficiency (in particular  for the O3
lines).  This might be caused by a decrease in the width of  
the $\ion{He}{ii}$
Ly-$\alpha$ profile 
together with a  decrease in the optical depth of the $\ion{O}{iii}$
resonance
lines.

\subsection{A detailed examination of the Bowen decays }

In this section all of the transitions that enter and exit each energy
level involved in the Bowen O1 fluorescence process are examined in
detail in order to verify the balance in the number of  photons 
(see Fig.6, Table 3, and Table 4).
 The lists of lines that
enter or exit the relevant levels and their A$_{ij}$ values  have been
obtained
 with the help of
the extensive compilation of levels and transitions by van Hoof in the
"The Atomic Line List 2.05" (http://www.pa.uky.edu/$\sim$peter/newpage).

\begin{enumerate} 
\item The 2p3d $^3P_2^o$ level (40.85 eV, 329469.80 cm$^{-1}$), is
pumped
directly by the $\ion{He}{ii}$ Ly-$\alpha$ $\lambda$ 303.782 line. The
list of
all primary decays from this level includes, (besides the two resonance
lines $\lambda$  303.800  and 303.622) the 6  lines
at $\lambda$$\lambda$
2807.90,  2818.66,   2836.28,  3132.79,
 3428.62, and  3444.06, and two additional lines (with rather
low
A$_{ij}$ values) at $\lambda$ (air) 2187.02 (I=0.40 10$^{-13}$ erg
cm$^{-2}$
s$^{-1}$ ) and at
$\lambda$ (air) 2197.48 (I=0.21 10$^{-13}$ erg cm$^{-2}$ s$^{-1}$)
that
correspond to
decays from 2p3d $^3P_2^o$ down to 2$p^4$ $^3P_2$ and 2$p^4$ $^3P_1$ at
35.18 and 35.21 eV, respectively. These two
lines, albeit weak, have been detected in STIS spectra. The $\lambda$
2187.70  line falls in the wing of the $\ion{He}{ii}$  Fowler
line $\lambda$ 2187.28   but is clearly present,
while the other line at $\lambda$ 2198.17 is weak but unblended (Fig
12).
 To the best of our knowledge they have never been
reported so far in any astronomical source.

For lines  that come from the same upper level k, the relative
photon numbers are proportional to the respective A$_{jk}$.
From the data in Table 4 we note that the observed photon 
number for the six lines of the primary decay (see also Fig. 6) are in
good 
agreement with the the A$_{jk}$ transition rates .  

\begin{figure}
\centering
\resizebox{\hsize}{!}{\includegraphics{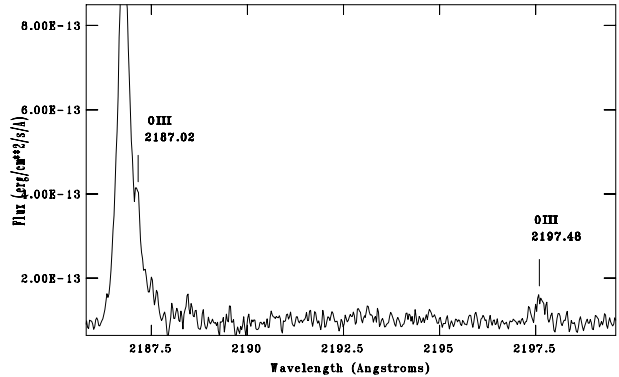}}
\caption{ Two "new"  Bowen lines from the  primary decay (40.85 eV).
The $\lambda$ 2187.02  line falls on the wings of the $\ion{He}{ii}$ 
Fowler
line $\lambda$ 2186.60, while the  $\lambda$ 2197.48  line 
is weak but unblended. All wavelengths are in air.}
\end{figure}

\item The 2p3p$^3P_2$ (300442.55 cm$^{-1}$, 37.25 eV) and the
2p3p$^3P_1$
(300311.96 cm$^{-1}$, 37.23 eV) levels are populated by the $\lambda$
3444.06 
and
the $\lambda$ 3428.62 decays, respectively (from 2p3d $^3P^o_2$).

Decays from these two levels include the five observed lines at
$\lambda$$\lambda$
3023.45, 3047.13,
3024.57, 3035.43, and 3059.30 (total A$_{ij}$ = 3.99 10$^8$) and  five
EUV
lines  with  lower term  2s 2p$^3$ $^3$D$^o_{1,2,3}$  near $\lambda$
554.5
(total  A$_{ij}$ = 2.36 10$^8$), plus other
weaker decays.

From the relative A$_{ij}$ values one would expect that the contribution
by the 
EUV lines should be about 0.67   of that of the observed near-UV lines 
(in the number of decays).
The total number of decays  in the two parent lines ($\lambda$  3428.62 
and $\lambda$  3444.06) is = 640.0 10$^{-3}$ s$^{-1}$  (Table 4).  The
same
quantity for the
five subsequent lines  at $\lambda$$\lambda$ 3047.13, 3023.45, 3059.30,
 3035.43,
 3024.57  amounts to 337.1 10$^{-3}$ cm$^{-2}$ s$^{-1}$. Therefore there
is
an apparent  (moderate - about 12 $\%$) deficiency of decays in these
five
observed  transitions, if  the
A$_{ij}$ values are accurate.

Inspection of the other possible  transitions listed in the  "Atomic
Line
list 2.05"
shows that, besides some
forbidden
transitions, there are two additional decays from 2p3p$^3P_1$ and
2p3p$^3P_2$ down to  2s2p$^3$ $^3S^o_1$ (197087.70 cm$^{-1}$, 24.44 eV),
via $\lambda$
968.76 , and $\lambda$ 967.54 respectively (mult. UV 17.05 ), whose Aij
values are close to 1.24e+06, that is, about 1 $\%$ of those of the
five lines above. These lines are also reported in Table 7 of Saraph 
and Seaton (1980).
However, inspection of BEFS spectra taken in Nov. 1996 has not led to
their
detection.

\item The 2p3p$^3S_1$ level (297558.66 cm$^{-1}$, 36.89 eV).

This level is important because it is fed by $\lambda$ 3132.79, the
strongest primary line in the decay from 2p3d $^3P_2^o$ (the O1
process). It is fed also by $\lambda$ 3121.64, the strongest primary
line
in the
decay from 2p3d $^3P_1^o$ (O3 process) and by $\lambda$ 3115.68, the
strongest decay from 2p3d$^3P_0^o$, but the
contribution from this latter line is almost negligible.

From the data of Table 4 we note that there is a severe unbalance
between the total number of photons in the
$\lambda$$\lambda$ 3132.79 and
3121.64 lines (= 1674.0 10$^{-3}$ cm$^{-2}$ s$^{-1}$) and the
corresponding
quantity (= 515.7 10$^{-3}$ cm$^{-2}$ s$^{-1}$) for the three observed
subsequent decay transitions at $\lambda$$\lambda$ 3340.74, 3312.30, and
3299.36.

Inspection of  "The Atomic  Lines List 2.05" and of the relevant 
radiative
rates
(Aij)
shows that additional decays
from 2p3p$^3S_1$ that could contain  the missing photons pass through
2s2p$^3$ $^3P^o_j$ (J=0,1,2) (142381.0, 142381.8, and 142393.5
cm$^{-1}$,
$\sim$
17.65 eV) with three lines that fall close to $\lambda$  644.44  (mult.
UV 16.20).
The sum of the Aij values of these $\lambda$ 644.44 lines is larger by
a factor 2.5 relative to that of the three optical lines. In other
words, about 71 $\%$ of the decay photons go into the three lines 
near $\lambda$ 644, thus reconciling the inbalance in the number of
decays. The expected intensity of the $\lambda$ 644 lines with
respect to that ($\sim$ 31.0 10$^{-13}$ erg cm$^{-2}$ s$^{-1}$) of the
three lines near $\lambda$  3320 ($\lambda$$\lambda$ 3340.74, 3312.30,
and 3299.36) should be larger by a factor 12.9 
(the relative A$_{ij}$ ratio   $\sim$  2.5   multplied by the  factor
3320/644). Therefore, the $\lambda$ 644 lines should have a total 
intensity of
$\sim$ 400 10$^{-13}$ erg cm$^{-2}$ s$^{-1}$ and constitute an
important
EUV contributor to the Bowen decays.
The  $\lambda$ 644 lines  were reported in the spectrum of the Sun
obtained
by Behring (1976) in the  study by Bhatia et al. (1982) and in the
Skylab spectrum of the Sun obtained by
Vernazza and Reeves (1978).  Their observed  intensities  were found    
by Bhatia et al. (1982)  to be in excellent agreement with
those obtained in a detailed calculation of the $\ion{O}{iii}$ EUV
spectrum of the
quiet Sun, when the process of photoexcitation by $\ion{He}{ii}$
Ly-$\alpha$
was included. 

Moore's tables (1993) report also a transition (mult. UV 17.04) from
2p3p$^3S_1$ (297558.66 cm$^{-1}$, 36.89 eV)  to 2s 2p$^3$ $^3S^o_1$
 (197087.7 cm$^{-1}$, 24.44 eV) that corresponds to the  $\lambda$ 995.31 
 line. This transition
between two S levels violates the pure LS coupling selection rules and
the line intensity should be very weak. However, it is worth reporting
the presence on BEFS spectra of a medium-weak line at $\lambda$ 995.3 (I
$\sim$
1.9 10$^{-13}$ erg cm$^{-2}$
s$^{-1}$) that lacks any other reliable identification.
It is worth noting that the Froese Fisher atomic
model for $\ion{O}{iii}$ supports  intermediate coupling versus pure LS
coupling.   We thank the referee for pointing out  the possibility that 
the 2p3p$^3S_1$ level be mixed with level  2p3p$^3P_1$. This interaction
between
states of different L is an indication of departure from pure LS
coupling.

\item The three 2p3p $^3D_{1,2,3}$ levels (294223.07 cm$^{-1}$, 36.48
eV,
294002.86 cm$^{-1}$ 36.45 eV, and 293866.49 cm$^{-1}$ 36.43 eV) are fed
by three
decays from 2p3d $3P^o_2$ (329469.80 cm$^{-1}$, 40.85
eV) at  $\lambda$ 2836.28, $\lambda$ 2818.66 and $\lambda$ 2807.90
 respectively, 
and produce six rather weak lines that fall between $\lambda$
3754 and $\lambda$ 3810.

The decay from 2p3p $^3D_3$ (294223.07 cm$^{-1}$) produces the
$\lambda$ 3759.87 
line, the only secondary line that is a "pure" O1 line. Unfortunately
the line falls in a blend with the strong and wide [$\ion{Fe}{vii}$] line
$\lambda$
3759. This prevents a detailed analysis. Additional
decays (with slightly weaker Aij) from this level are possible through
the EUV lines at $\lambda$ 658.58 and $\lambda$ 574.06.

The observed decays from the 2p3p $^3D_2$ level (294002.86 cm$^{-1}$)
produce
the two weak lines at $\lambda$ 3754.67 and $\lambda$ 3791.26. The sum
of the
number of photons in these two  lines (= 21.4 10$^{-3}$ cm$^{-2}$
s$^{-1}$ ) is close to that (= 18.5 10$^{-3}$ cm$^{-2}$
s$^{-1}$) in the
parent line $\lambda$ 2818.66.

However, from  "The Atomic Line List  2.05" one can see  that there
are
competitive decays from 2p3p $^3D_2$
with similar Aij in the EUV region at $\lambda$ 574.8, and $\lambda$  
659.5.
Therefore, either the $\lambda$  3754.67 and $\lambda$  3791.26 lines
are
slightly
blended or the 2p3p $^3D_2$ level is partially populated by the
charge-exchange  (CE) mechanism,  as reported in  the following lines.

The decays from level 2p3p $^3D_1$ (36.43 eV, 293866.49 cm$^{-1}$)
produce the three lines at
$\lambda$$\lambda$ 3757.21, 3774.00, and 3810.96. This level is the
lower
level of the $\lambda$
2807.90 line. It is worth noting  that the $\lambda$  2807.90 line is
barely
detectable (N$_{\nu}$ = 0.4 10$^{-3}$ cm$^{-2}$ s$^{-1}$ ) in the STIS
near UV
spectrum that has  high resolution  (see Fig. 7); instead, two of the
subsequent decay
lines,
 i.e.  $\lambda$ 3757.21 and $\lambda$ 3774.40, are present  both in
STIS and UVES spectra  (N = 9.96
10$^{-3}$ cm$^{-2}$ s$^{-1}$ ) while the third
one
at $\lambda$ 3810.96 falls in a blend. This suggests that the 2p3p
$^3D_1$
level is mainly populated by charge-exchange, as reported in Aller
(1984) and
Dalgarno and Sternberg (1982).

\item

The three 2p3s$^3P^o_{0,1,2}$ levels (267634.00, 267377.11, and
267258.71 cm$^{-1}$, 33.18, 33.15, and 33.14 eV) are populated
by
14 near UV/optical lines of the secondary decays from levels 2p3p
$^3P_{0,1,2}$, 2p3p $^3S_1$, and 2p3p $^3D_{1,2,3}$. The  number
of decays to these levels (with a negligible contribution from the CE
process) is 516.3, 259.1 and 102.5  10$^{-3}$ cm$^{-2}$ s$^{-1}$
respectively. The total number of decays is  878 10$^{-3}$ cm$^{-2}$
s$^{-1}$ to be compared
with the
2444.1 10$^{-3}$ cm$^{-2}$ s$^{-1}$ primary decays from level 2p3d
$^3P_2^o$ (O1).
We note that the sole permitted decays from these three
2p3s$^3P^o_{0,1,2}$ levels are
through the six
EUV resonance lines of mult. UV 6 near $\lambda$ 374 (see also Fig. 6) 
that are important in the context  of the Bowen secondary $\ion{N}{iii}$
fluorescence (see Sect. 6.1).

\end{enumerate}

We point out that unlike the primary lines that are "pure" O1 or O3
lines,
most secondary lines are "mixed" O1 + O3 lines (see also sections 4.1 and
4.4).
 However, most O3 lines
are very weak, and their contribution to the secondary decays is
generally
negligible. The sole exception is that  of the $\lambda$ 3122
line which, however, represents a contribution
of about  2$\%$  to the total population of the 2p3p $^3S_1$  
level.
This justifies the neglect of the
O3 contribution to the secondary lines in most of the previous
considerations.

\section{The other  $\ion{O}{iii}$ lines}

In principle, recombinations and charge-exchange (CE) could be effective
in
populating
some high levels of $\ion{O}{iii}$ and provide some contribution to the
intensity of the Bowen lines.

To check for the effectiveness and the relative importance of these
processes we have looked for the presence in STIS and UVES spectra of
such $\ion{O}{iii}$ lines coming from high-lying levels, whose
excitation is
comparable to that of the Bowen lines.

The inspection has revealed the presence of a small number of weak lines
in
the near-UV optical range, the strongest one being the $\lambda$
2983.81 line (mult. UV 18,  u.l. = 38.01 eV, I = 9.2 10$^{-14}$ erg
cm$^{-2}$ s$^{-1}$). The only other
possible detection in the near-UV is the (weak) line at $\lambda$
2959.72
(UV 19.12, 40.26 eV)  with  I= 1.44 10$^{-14}$ erg cm$^{-2}$ s$^{-1}$).
This line would feed the 
$\lambda$ 5592.25  line (mult 5,  u.l. = 36.07 eV), that is important in
the
context of the charge-exchange process (see below). 

In the optical range, in UVES spectra, we have detected the three weak
lines of mult. 8 (u.l. = 40.27 eV) at $\lambda$ 3265.32 (I=2.54
10$^{-14}$
erg cm$^{-2}$ s$^{-1}$), 
$\lambda$ 3260.84 (I=1.76 10$^{-14}$ erg cm$^{-2}$ s$^{-1}$) and 
$\lambda$ 3267.20 (I=2.10 10$^{-14}$ erg cm$^{-2}$ s$^{-1}$), 
 and the $\lambda$ 3455.06 line of mult. 25 (u.l. 
= 48.95
eV) with  I = 0.96 10$^{-14}$ erg cm$^{-2}$ s$^{-1}$). The important
$\lambda$
5592.25  line of mult. 5,  (u.l.=36.07 eV)
whose presence is associated with the CE process  is detected only
in
UVES, with intensity of 1.2 10$^{-14}$ erg cm$^{-2}$ s$^{-1}$.

Therefore, the contribution by recombinations and/or CE to the
observed
intensity  of the $\ion{O}{iii}$
Bowen lines is generally negligible,  with the exception of  the six weak 
lines in the decay from the 2p3p$^3D_{1,2,3}$
term, whose individual intensities are in the range of 0.2 - 0.8
$10^{-13}$erg cm$^{-2}$ s$^{-1}$.

\section{The $\ion{N}{iii}$ 4640 lines}

\subsection{The excitation mechanism}

Bowen (1934, 1935) pointed out that, by another remarkable coincidence
in nature, the $\ion{O}{iii}$ resonance line $\lambda$  374.432 (one of
the six
decays from 2p3s $^3P^o_{0,1,2}$ to the ground term 2$p^2$ $^3P_{0,1,2}$
in the
final decays of the Bowen lines) has nearly the same wavelength
\footnote {  Wavelengths for the $\ion{N}{iii}$ and $\ion{O}{iii}$ lines
in
this Section and in the following  are from "The 
Atomic Lines List 2.05" (http://www.pa.uky.edu/$\sim$peter/newpage),
rounded to
the third decimal digit. We note
that these wavelengths are the same as those reported in Wiese et al.
(1996),
and in the NIST Web site 
(http://physics.nist.gov/PhysRefData/ASD/index.html) as Ritz wavelengths.
 These wavelengths come from 
the analysis by Pettersson, 1982 and they are supposed to be correct to 
about 0.0005 \AA~ at 250 \AA~ and to 0.002 \AA~ at 500 \AA. 
Interpolation 
suggests a wavelength uncertainty of 0.00125 \AA~ (1.0 km s$^{-1}$)
 near $\lambda$  374).} 
as the two resonance lines of $\ion{N}{iii}$ $\lambda$ 374.434 and
$\lambda$  374.442. Therefore, photoexcitation by 
$\ion{O}{iii}$  374.432
can populate both the 3d $^2D_{5/2}$ level (267244 cm$^{-1}$) and the 3d
$^2D_{3/2}$ level (267238.40 cm$^{-1}$) of
$\ion{N}{iii}$ from the  2p $^2P^o_{3/2}$ level (174 cm$^{-1}$) of the
ground
term. In
the decay from these two high levels, the three lines at
$\lambda$$\lambda$
4640.64, 4641.85 and  4634.13 are emitted, and in the subsequent decay
two additional lines at $\lambda$  4097.36 and $\lambda$  4103.39 are
produced. The relevant levels and transitions are reported in Fig. 13
and Table 7.   Each level has also been identified by a number index from
1
to 7 
in order to facilitate reading the text.
See also Kastner and Bhatia (1991) (hereinafter KB91)  and Kallman and
McCray 
(1980) for  further theoretical
considerations
and quantitative evaluations.

These optical $\ion{N}{iii}$ lines are observed as quite strong emission
lines in
planetary nebulae, x-ray binaries, symbiotic stars and novae in the
early nebular stages.

KB91, however, from a direct
comparison of the
observed line ratios with the theoretically predicted ratios expected
from the postulated Bowen process of selective photoexcitation,  have
challenged the common interpretation that these $\ion{N}{iii}$ emission
lines 
originate from a secondary Bowen
fluorescence. Their main argument against Bowen fluorescence is that
the observed I$_{4634}$/I$_{4640}$ ratio indicates a relative population
ratio
N$_6$/N$_7$  close to the "statistical"
value 0.667   (where  level 6 is
3d $^2D_{3/2}$ and  level 7 is 3d $^2D_{5/2}$, see also Fig. 13).
Instead, in the case of
Bowen fluorescence,  a  much lower value (about 1/9) is expected as a
consequence of the fact that A$_{2,19}$/A$_{2,20}$ $\sim$ 0.167 and
g$_{19}$/g$_{20}$=4/6.  KB91 did not indicate which physical process
could be responsible for the thermal population ratio that apparently
existed. After tentatively suggesting recombination and
charge-exchange, they ruled out both processes  after specific
considerations.

In a subsequent paper, Ferland (1992) suggested that the optical
$\ion{N}{iii}$
lines could be excited by direct continuum fluorescence (CF). While
the
Bowen mechanism would pump both the 3d $^2D_{3/2}$ and 3d $^2D_{5/2}$
levels from
the excited level 2p $^2P_{3/2}$ of the ground term (with the A$_{ij}$
and g
factors favoring level 3d $^2D_{5/2}$ over level 3d $^2D_{3/2}$)
continuum
fluorescence would pump the  level 3d $^2D_{5/2}$ from the excited level
of
the ground term and the 3d $^2D_{3/2}$ level predominantly from the
ground
level 2p $^2P_{1/2}$ of the ground term, (via the  374.198 transition)
with a
minor contribution (about 20 $\%$) from the excited level. As a result,
 (the A$_{ij}$ of the transition from the $^2P_{1/2}$ level of the ground
term
being comparable to that of the transition from the $^2P_{3/2}$ level)
the
relative population of the 3d $^2D_{5/2}$ and 3d $^2D_{3/2}$ levels
would
become
comparable, with the same result for the intensities of the three decay
lines near $\lambda$ 4640, thus reconciling the predicted intensities
with the observations.

\begin{figure}
\centering
\resizebox{\hsize}{!}{\includegraphics{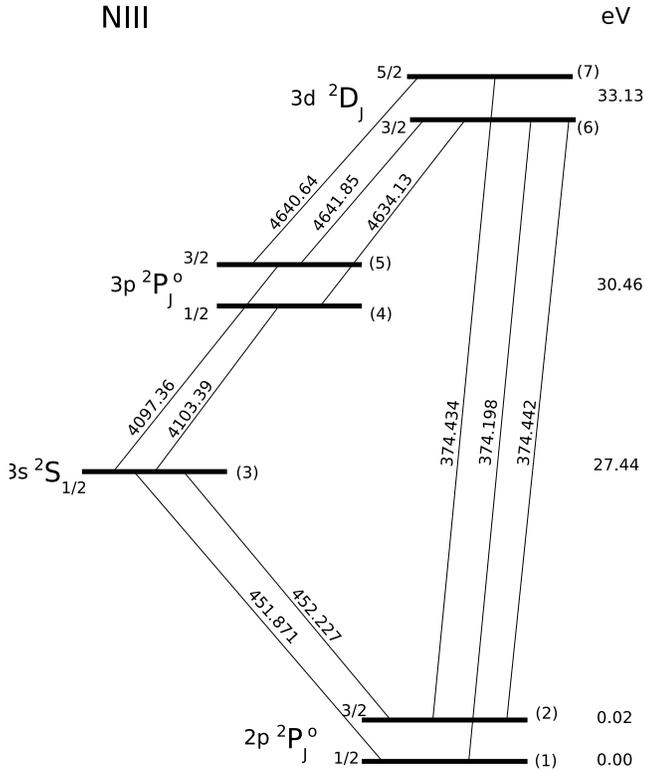}}
\caption{ A partial Grotrian diagram for $\ion{N}{iii}$.  Levels are
identified by a number index 1-7 to facilitate  reading the text.
 }
 \end{figure}

In a very recent paper, Eriksson et al. (2005)  have presented a set of
semi-empirical equations for the prediction of the relative intensities
for the 
$\ion{N}{iii}$ lines that are generated by the Bowen mechanism. They
have  also
suggested an additional
pumping
channel associated with the $\ion{O}{iii}$ $\lambda$  374.162 line (one
of
the six
resonance
lines in the final decay of the primary Bowen mechanism) that could
pump the $\ion{N}{iii}$  $\lambda$  374.198 line.

Eriksson et al. have obtained the intensities of the $\ion{O}{iii}$
Bowen excited
lines (2800-3900 \AA) from IUE data of 1993 and ground-based data of
the same year (Mc Kenna et al. 1997). The $\ion{N}{iii}$ line
intensities  have
been obtained also from Mc Kenna et al.(1997). Instead, the line
widths for the $\ion{O}{iii}$ and $\ion{N}{iii}$ lines  and the relative
velocity shifts
have been obtained from IUE data (SWP29535) of 1986 (see  section 6.4
for further comments).

In the case of RR Tel, Eriksson et al. have predicted a line ratio
(relative strength) 
I$_{4634}$
/I$_{4640}$ = 0.245, which is too low relative to their observed values
(=
0.47), and  have ruled out this additional line fluorescence channel
as the main one responsible for populating level 6. They have concluded
that the two 3d $^2D_{3/2,5/2}$ levels are predominantly populated by
processes other than the Bowen mechanism and have  suggested that
radiative recombination could be the
main population process of these levels, although,
as they
pointed out, this process cannot explain some discrepancies between the
predicted and observed relative intensities of the 4641.85 and 4640.64
lines.

\begin{table}
\caption{Wavelengths, A$_{ij}$  values, STIS and FEROS 
intensities
($10^{-13}$erg 
cm$^{-2}$s$^{-1}$), FWHM (km s$^{-1}$),  
relative intensities (I$_{4640.64}$ = 1.00), and photon number for the
$\ion{N}{iii}$  emission
lines 
associated with  the  Bowen fluorescence. The  continuum  of FEROS has
been
scaled  to that of STIS.}
\begin{center}
\renewcommand{\tabcolsep}{0.1cm}
\begin{tabular}{ccccccc}
\hline
 $\lambda_{air}(\AA)$ & A$_{ij}$  &  STIS & FEROS   & FWHM &  Rel. Int.&
N$_{\nu}$\\                              
 4097.36& 8.66e+07&   2.05  &1.75  &  35.3    & 0.59  &  0.042\\
 4103.39& 8.62e+07&    -    & (0.90)  & (31.0)  & (0.31)  & (0.022) \\
 4634.13 & 6.00e+07&   1.69   & 1.41 &  32.0    & 0.49 &  0.039\\
4640.64 &  7.17e+07&    3.46  & 2.89 &  31.8       & 1.00 & 0.081   \\
 4641.85 & 1.19e+07&  (0.60)  & 0.40 & 33.7       & 0.14  & (0.014) \\
 374.198 & 1.05e+10 &   -     & -  &-&-& -\\
374.434 & 1.26e+10&   -    &  -  &-&-& -\\
374.442 & 0.21e+10&   -    &   -  &  - & -&- \\

\hline
\end{tabular}
\end{center}
\end{table}

\subsection {The  $\ion{N}{iii}$ lines intensities in RR Tel}

ELI and FWHMs for all $\ion{N}{iii}$ subordinate lines
specifically involved in the Bowen fluorescence process have been
measured in STIS and FEROS spectra.

We note (see also section 2.1) that the STIS data are absolutely
calibrated but suffer from a
limited spectral resolution in the optical range (about 8000). Instead,
the FEROS data are not absolutely calibrated but have higher spectral
resolution (about 60,000). Thus, STIS data provide a quite good estimate
of the absolute line flux,
while the FEROS data provide very accurate line ratios and good FWHMs
measurements.

The re-calibration of FEROS (see section 2.4)  has also allowed us to
obtain reliable ELI
and reliable line ratios
for a few
 $\ion{N}{iii}$  emission lines that are clearly observed on FEROS but
are not
detected or are confused with noise in the
STIS grating spectra in the optical range.

 Table 7 gives
the measurements
for the individual lines and their average (STIS and FEROS) intensities
relative to the reference line $\lambda$ 4640.64.
The  intensity of the   4103.39 line is rather uncertain because it falls
on
the red wing of the strong H$_{\gamma}$ line.
The data also indicate
that
I$_{4640.64}$/(I$_{ 4641.85}$+I$_{ 4634.13}$) = 1.59  and
I$_{ 4634.13}$/I$_{4640.64}$ = 0.49, close to the value (0.47)
found by Eriksson et al. (2005).

The $\ion{N}{iii}$ lines have FWHM values near 33.2 km s$^{-1}$, a value
that
is close to  that (35.3 km s$^{-1}$) found for the $\ion{O}{iii}$ Bowen
lines. The
average relative displacement ($\ion{O}{iii}$ - $\ion{N}{iii}$) is 
-1.45 km s$^{-1}$ 
(Eriksson
et al. instead found  4-5 km/s).

\begin{figure}
\centering
\resizebox{\hsize}{!}{\includegraphics{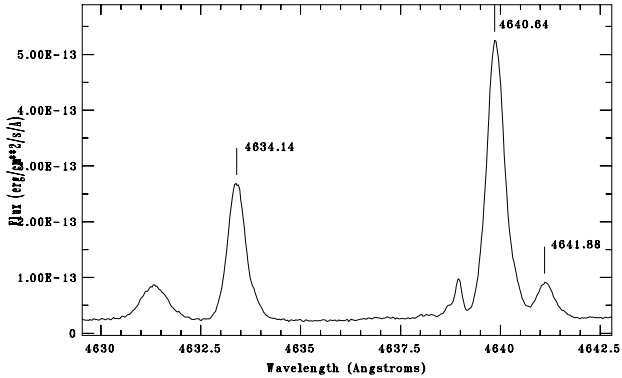}}
\caption{  The FEROS spectrum of the  three $\ion{N}{iii}$  lines  close
to
$\lambda$ 4640. The continuum has been scaled to that of STIS.}
 \end{figure}

\subsection {The exclusion of continuum fluorescence and radiative
recombination
as the   excitation  mechanism of the $\ion{N}{iii}$ $\lambda$
4640 lines.}

As already mentioned in section 6.1, in order to explain the fact that
the the $\ion{N}{iii}$
 $\lambda$ 4640 lines have relative intensities that are indicative of a
"statistical"
population of the 3d $^2D_{J}$  levels, Ferland (1992) argued that these
lines
are the result of continuum fluorescence. 

However, if this mechanism were effective for the three $\ion{N}{iii}$
lines near
$\lambda$ 374 (mult. UV 5) one would expect a similar excitation
mechanism
also for the other EUV resonance lines of $\ion{N}{iii}$. Inspection of 
"The Atomic Line List  2.05"  and of the Moore's (1993)
tables of spectra  in the range $\lambda$ $\lambda$ 250-500 shows
the
presence of several resonance lines (multiplets) with Aij values
(intensities) similar to those of the 374 lines, i.e. 
$\lambda$$\lambda$
451.87 + 452.23.
(UV 4), 332.33 + 332.14 (UV 5.01), 323.49 + 323.61 + 323.67 (UV 6),
314.71
+ 314.86 + 314.89 (UV 7) , 311.64 + 311.55 + 311.72 (UV 7.01), 292.44 +
292.59 (UV 7.04), 282.21 + 282.07 (UV 7.06) and  other multiplets of
weaker
intensity. For each level of the upper term of these multiplets we have
selected the strongest decays (about a dozen lines
between
$\lambda$ 1385 and $\lambda$ 4205, i.e. $\lambda$$\lambda$ 1387.30,
1387.38, 1387.99,
1804.49,
 1885.06, 1885.22, 2248.36, 2249.63, 2334.26, 2335.61, 3304.98,
3307.58, 3307.70, 4201.26, 4855.19, 4874.46, 4882.03) and checked for
their
presence in the STIS + UVES + FEROS spectrum of RR Tel.

The search has given a definite negative result.
The absence of decays from the upper level of
strong resonance lines that are likely to be pumped by continuum
fluorescence is hardly compatible with the process
of CF since it is hard to explain how this mechanism might work only for
a selected group
of lines viz. the 4640s.

Moreover, if CF were effective, it should work also in the case of the
$\ion{O}{iii}$ resonance lines (and other resonance lines in that range)
and one
would expect, for example,  CF to pump the three upper
levels $(2p 3d)^3P_{2,1,0}$ of the six resonance lines near $\lambda$
303. In this case, from the  A$_{ij}$ and g values (the I$_{ij}$ values
being the same for all transitions) one would expect to see a
"statistical"
relative  population ratio 5:3:1 within these three upper levels and
therefore a
corresponding relative intensity ratio for the three "pure" primary
lines ($\lambda$
3132.79, $\lambda$ 3121.64 and $\lambda$ 3115.68) that emanate from these
levels
(since they all have similar A$_{ij}$ values = 1.50 10$^8$), unlike what
is
observed. If, instead,  one assumes that the observed intensity is a
combination of contributions from Bowen fluorescence  and continuum
fluorescence
this would imply that CF, if present, is, at best, very marginal since
observations show that the $(2p 3d)^3P_0$ level (upper level of
$\lambda$ 3115.68 ) is very weakly populated.

Concerning other physical processes such as radiative recombination as
the 
main population process of the 3d $^2D_{J}$ levels (see also the
considerations by KB91 and Eriksson et al. 2005 in section 6.1), we
have checked in detail the possible presence of recombination
lines of $\ion{N}{iii}$ in the RR Tel spectrum.
In particular we have searched in  "The Atomic Lines List 2.05" all
possible
decays
into the two upper levels 3d $^2D_{5/2}$ ( 267244 cm$^{-1}$) and 3d
$^2D_{3/2}$ (267238.40 cm$^{-1}$) of the $\ion{N}{iii}$ 4640 lines, that
is for 
possible
parents of these lines. The strongest lines, besides a sextet near
$\lambda$ 1324,
are
those of mult. UV 23 with three lines near $\lambda$ 2248 and those of
mult. UV 24 with two lines near $\lambda$ 1885. None of these lines has
been detected. Therefore the  4640 lines are not fed
by decays from higher levels. We have also checked for the presence of
high-lying subordinate lines with high Aij and excitation level similar
to that of the 4640 lines, e.g. the lines of mult. UV 22, 22.01, and
22.02., but  this search has also given negative results.

An inspection of the RR Tel spectrum for the possible presence of the
strongest transitions into the two lower levels 3p $^2P^o_{J}$ (17 and
18) of the three 4640 lines, e.g. $\lambda$
1387.30,  $\lambda$  1387.38,  and $\lambda $1804.49 (whose sum of the
A$_{ij}$ is about ten
times larger than the A$_{ij}$ of the $\lambda$  4634.13 and $\lambda$ 
 4641.85 lines) has also given a
negative result.

Therefore, recombination is ruled out by the absence of possible decays
from higher levels into the
three 4640  lines, and the absence of other subordinate lines with
similar
Aij and excitation as the 4640 lines.
In conclusion:  neither continuum fluorescence  nor radiative
recombination
 seem effective in producing  the
observed  $\ion{N}{iii}$ emission lines in RR Tel. It is hard to explain
how
these 
processes may  excite levels 6 and 7 of $\ion{N}{iii}$ only, which
indicates,  instead, that some kind of selective  process is
present.

\subsection {The role of multiple  scatterings in the resonance lines of
$\ion{O}{iii}$ and $\ion{N}{iii}$}

Continuum fluorescence and radiative recombination being ruled out,  we
must consider
if and how a selective process like line
fluorescence, could be responsible for the observed 
 relative  lines ratio in the $\ion{N}{iii}$ 4640 lines.

We are supported in this investigation  by the circumstantial evidence
that the
presence of the $\ion{O}{iii}$ Bowen lines is generally associated with
that of
the $\ion{N}{iii}$ 4640 lines in all well studied objects, a clear
indication
that the  excitation mechanism is similar.

We recall that Eriksson et al. (2005) have taken into account the
possibility of an extra pumping of level 6 (3d $^2D_{3/2}$) from
level 1
(which is required for approaching the observed I$_{4634}$/I$_{4640}$
line ratio)
by the near coincidence between the $\ion{N}{iii}$ $\lambda$  374.198 and
the $\ion{O}{iii}$
$\lambda$  374.162
lines.  They have obtained a population ratio N$_6$/N$_7$ = 0.29
and a line
ratio I$_{4634}$/I$_{4640}$ = 0.245 which is too low relative to their
observed values
(= 0.47) and concluded that this mechanism is not effective.

 Eriksson et al. (2005), using the study of Pettersson (1982) 
  have also  pointed out that the combined effects
 of the  uncertainties 
in the wavelengths of the ground term transitions of $\ion{N}{iii}$
and $\ion{O}{iii}$ near  $\lambda$  374  and of the 
relative velocity
shifts could introduce  larger uncertainties for the overlap of the 
profiles 
of the pumping and of the pumped lines  and therefore
for the relative line strengths. However, the Pettersson paper (1982)
suggests
 uncertainties of 0.0005 \AA~ at 250 \AA~ and 0.002 \AA~ at 500 \AA~,
which
 means an uncertainty of 1.0 km s$^{-1}$.

The velocity separation between $\ion{O}{iii}$ $\lambda$  374.162 and
$\ion{N}{iii}$ $\lambda$  374.198
is close to
30
kms$^{-1}$ and the effectiveness of pumping of $\ion{N}{iii}$ level 6
from level
1  by 
$\ion{O}{iii}$ $\lambda$  374.162  crucially
depends on the overlap between the two lines (instead, the
pumping of  levels 6 and 7 of $\ion{N}{iii}$ from level 2  are
easily explained by the almost full overlap (coincidence) between
the $\ion{O}{iii}$ $\lambda$  374.432 and the $\ion{N}{iii}$ $\lambda$
374.434 (N1) and
$\lambda$  374.442 (N2) lines).
The negative conclusion by Eriksson et al. (2005) derives  from  the 
(allegedly) small overlap
between
the
profiles of the  $\ion{O}{iii}$ $\lambda$  374.162 and $\ion{N}{iii}$
$\lambda$  374.198  lines,
that could not
guarantee the
required pumping of level 6 from level 1.

In this context, it should be noted that Eriksson et al have based their
calculations of line overlap on the measurements of the gaussian widths
(=0.6*FWHM) of the {\em intercombination} lines of $\ion{N}{iii}$] 
$\lambda$ 1750
and $\ion{O}{iii}$] $\lambda$ 1660
 in IUE spectra of 1986. The same width has been assumed in the
calculation of
the overlap between the profiles of the {\em resonance} lines of 
 $\ion{O}{iii}$
$\lambda$ 374.162 and $\ion{N}{iii}$
 $\lambda$ 374.198. 
Instead, in STIS and FEROS spectra, the subordinate lines of $\ion{O}
{iii}$ and
$\ion{N}{iii}$ are wider than the intercombination lines by about 5 km
s$^{-1}$,
with 
the effect of increasing  the line overlap.

However, what is of much higher relevance
is the fact  that  both the six $\ion{O}{iii}$ $\lambda$ 374
resonance lines that connect the ground term 2p$^2$ $^3$P with the term
2p3s
$^3$P and  the three $\ion{N}{iii}$ resonance lines $\lambda$ 374 that
connect
levels 1
and 2 with levels 6 and 7 are all {\em optically thick}, with optical
depths on the order of  1000 (see appendix for an approximate estimate
), while Eriksson et al. (2005) in their semi-empirical calculations
have
assumed that all transitions are optically thin.

We suggest here that  multiple scattering in the $\ion{O}{iii}$
$\lambda$  374.162
resonance  line
 has the effect of increasing the pumping of level 6 from level 1  of
$\ion{N}{iii}$, 
as compared to the optically thin case.
Under optically thick conditions the $\ion{O}{iii}$ resonance-line
photons will
suffer many scatterings ($<$N$>$ $\sim$
$\tau$$_o$$\cdot$$\sqrt{\ln(\tau_o)}$) within
the nebula. The probability per single scattering that a $\lambda$
 374.162 
photon of $\ion{O}{iii}$ will excite level 6 of $\ion{N}{iii}$ (instead
of level 2p3s
$^3P_1$ of
$\ion{O}{iii}$)
depends mainly on the overlap between the $\ion{O}{iii}$ $\lambda$
 374.162 and the
$\ion{N}{iii}$ $\lambda$  374.198
lines and on the relative $\ion{N}{iii}$/$\ion{O}{iii}$ ground term
population (abundances-
ionization fraction) ratio.

We have no direct information on the widths of the $\ion{O}{iii}$
resonance lines.
In our spectra the observed subordinate lines (most of them being the
fluorescence lines) have FWHM close to 35.3 km s$^{-1}$ (while the
intercombination lines of $\ion{O}{iii}$ near $\lambda$ 1666  and  those
of $\ion{N}{iii}$
lines
near $\lambda$ 1750  have
smaller FWHMs values,  close to 29.7 and 25.7 km s$^{-1}$ 
respectively).
Assuming gaussian widths, the line overlap fraction is close to 0.30.
However, this is almost certainly a lower limit since it is likely that
the
resonance lines are wider than the recombination and the 
intercombination
lines because of line broadening by Doppler shifts. Opacity broadening
($\propto$  $\sqrt{\ln(\tau_o)}$)  will
significantly affect their shape and line center optical depths close
to 1000 will result in line widths 2.6 times larger than that of the
optically thin case (Oegerle et al. 1983). Also, the shape of
the resonance lines is better represented by a Lorentzian profile, with
wider wings than the Gaussian one. Moreover as mentioned in sect. 4.3,
the profile of the $\ion{O}{iii}$ Bowen line is aymmetric toward the red
with a
maximum velocity in the red wing of about 100 kms$^{-1}$.

In any case, with the assumptions that : 1) the overlap between the
$\ion{O}{iii}$
$\lambda$  374.162 and the $\ion{N}{iii}$ $\lambda$  374.198 resonance
lines
is about
0.30 (a lower limit, as derived from the subordinate lines of the same
ions), 2) the abundance ratio $\ion{N}{iii}$/$\ion{O}{iii}$ is about
1/15 (the
abundance ratio N/O is about 3 and the ionization fraction ratio
($\ion{N}{iii}$/N)/($\ion{O}{iii}$/O) is about 5 in the region where
the two ions
co-exist, HN86),   3) all
$\ion{N}{iii}$
 is in the ground term (with  relative population of the 2p
$^2P^o_{1/2}$ 
level given by its statistical weight  = 1/3), one obtains a
rough estimate of 0.30 $\cdot$ 0.0667 $\cdot$ 0.333= 6.67 10$^{-3}$ as
the
direct probability per single  scattering that a $\lambda$  374.162
photon
of $\ion{O}{iii}$ will directly excite level 6 of $\ion{N}{iii}$ from
its ground level
$^2P^o_{1/2}$ (Hydrogen photoionization is neglected: He$^{+2}$ region:
no neutral H). However, since the optical depth (or the number of
scatterings before the line escapes the nebula) for the $\ion{O}{iii}$
resonance
$\lambda$  374.162 line is about 1000, (see appendix) the probability
that the level 6 of $\ion{N}{iii}$ is {\em not} excited after this
number of
scatterings is (1-0.0066)$^{1000}$ $\sim$ 0.001. See Ferland (1992) for
similar considerations in the case of resonance
fluorescence.

Once the 3d $^2D_{3/2}$ level (level 6) of $\ion{N}{iii}$ is pumped, it
can
decay either via
resonant scattering with emission of the (19$\to$1) $\lambda$  374.198
line
(or the
less likely (19$\to$2) $\lambda$  374.442 line) or via emission of one of
the
two
subordinate lines at $\lambda$  4634.13  and $\lambda$  4641.85.
The A$_{ij}$ branching ratio from the upper level (6) (see Table 7)
favours  resonant
scattering with respect to
the cascade routes via the $\lambda$  4641.85 and $\lambda$  4634.13 
lines  by a factor
about 175 = (1.26 10$^{10}$)/(7.19 10$^7$). Therefore, the probability
of conversion into the cascade route in a single act (decay) is =
0.0057.
However, if the number of scatterings of the $\ion{N}{iii}$ $\lambda$
 374.198
resonance line is about 1000  (see
Appendix), the
probability that the line will not be converted into the optical
$\lambda$ 4634 +
$\lambda$ 4641 lines  after this number of scatterings is on the order
(1 - (A$_{19,18}$ +
A$_{19,17}$)/(A$_{19,1}$+A$_{19,2}$))$^{1000}$   $\sim$
0.01.

Therefore,  multiple scattering of the  $\lambda$ 374  resonance lines
of $\ion{O}{iii}$
and
$\ion{N}{iii}$ have the net effect of converting a significant fraction
of these
 photons into
photons of the $\ion{N}{iii}$ subordinate lines at $\lambda$ 4634 and
$\lambda$
4641 
that can
easily escape from the nebula.   We point out that while the $\ion{O}
{iii}$ 
$\lambda$
374
lines are pure resonance lines,
without alternative decay routes from their upper levels,  the $\ion{N}
{iii}$
374 photons have an
escape route through conversion to  the  $\lambda$ $\sim$ 4640  and
$\lambda$ $\sim$ 4100 lines (and other decays).
 It is obvious  that  in the optically thin case the pumping
efficiency
is critically dependent  on the degree of line overlap, while in
the optically thick case  line broadening and multiple scattering will 
strongly increase the pumping efficiency.

It should  also be noted  that high oxygen and nitrogen abundances
(about 4  $\times$ solar), such 
as those generally found in the ejecta of novae would produce more
intense $\ion{O}{iii}$ Bowen lines as well as more intense $\ion{N}
{iii}$ $\lambda$ 4640
lines
because of the larger optical depths in the $\ion{O}{iii}$ and $\ion{N}
{iii}$ resonance
lines (see also Netzer et al. 1985).

Clearly, the extra contribution to the population of level 6 is
added to that produced by the excitation via the $\ion{O}{iii}$
$\lambda$ 
 374.432 line. Note  that the $\lambda$  374.162 and the 
$\lambda$ 374.432 $\ion{O}{iii}$ lines emanate
from the same upper level (see Fig. 6) and that the intensity ratio of
the two lines
is 
$\sim$ 0.600  because of the relative  A$_{ij}$ ratio (Table 8).
As already
mentioned in section 6.1, if level 6 is excited by  $\lambda$  374.442 
only, the population ratio N$_6$/N$_7$ is = 0.111 (because of the
relative Aij ratio = 0.167 and statistical weights ratios = 0.666) and
the expected
I$_{4634}$/I$_{4640}$
ratio becomes 
$\sim$ 0.09, while the observed ratio is $\sim$ 0.49. Therefore, 
in order to achieve this  line ratio the relative population ratio 
N$_6$/N$_7$ must be
close to about 0.57.

Simple  calculations of the photoexcitation rate of  
levels 6 and 7  of $\ion{N}{iii}$ from the ground term   ($\propto$
n$_i$
$\cdot$ g$_j$/g$_i$ $\cdot$ A$_{i,j}$ $\cdot$ I$_{i,j}$),  assuming LTE
population for the $\ion{N}{iii}$ ground 
levels, show that if all  $\ion{O}{iii}$  374.162 photons are effective 
in the
population of  level 6 of $\ion{N}{iii}$, the  
N$_6$/N$_7$ ratio becomes $\sim$ 0.444 and, correspondingly, the
I$_{4634}$/I$_{4640}$
ratio becomes $\sim $ 0.37. This ratio is slightly lower than the
observed
one  ($\sim$ 0.49).

\subsection{An additional pumping channel ?}

Without invoking deviations from LTE in the population of the two levels
of the ground term of $\ion{N}{iii}$ in order to obtain the required
ratio
(a
physical condition 
that would not be too surprising, on
account of the low electron density and of the fact that the radiative
processes seem to dominate over
the
collisional ones), we suggest here that the missing contribution 
 could arise also from the $\ion{O}{iii}$ $\lambda$ 374.073 line, one
of the two
decays from level 2p3s $^3P_2^o$. Although the line center separation
is 101 km s$^{-1}$, even a 
small overlap between the line wings  
would ensure some degree of pumping if the line optical depth is high.

Using the same considerations adopted in the previous section for the
$\ion{O}{iii}$
$\lambda$  374.162 line (but assuming, in  a conservative approach,
only 5  $\%$ as the
overlap
fraction between the  $\ion{N}{iii}$ $\lambda$  374.198 and the $\ion{O}
{iii}$ $\lambda$ 
 374.073
line profiles) one
obtains a rough estimate of 0.05 $\cdot $0.0667 $\cdot$ 0.333=
1.1 10$^{-3}$ as the direct probability per single scattering that a
$\ion{O}{iii}$ $\lambda$  374.073 photon   will directly excite level 6
of $\ion{N}{iii}$
from its
ground level $^2P_{1/2}$.
However, since the optical depth of the $\ion{O}{iii}$ $\lambda$  374.073 
resonance
line  is 
about 6000  (larger than that of the  $\ion{O}{iii}$ $\lambda$ 
 374.162 line),  
the probability that  level 6  of $\ion{N}{iii}$ is {\em not}
excited  after this number of scatterings is  again very low 
(1-0.0011)$^{6000}$ $\sim$ 0.0005.

One should also note that the smaller  overlap is compensated
by the much larger line intensity of $\ion{O}{iii}$ $\lambda$  374.073
 relative
to $\ion{O}{iii}$ $\lambda$  374.162.
In fact, as a result of the combined effect of the larger number of
decays 
to the $\ion{O}{iii}$
3p3s
$^3P^o_2$ level at 267634
cm$^{-1}$  relative to the $\ion{O}{iii}$ 3p3s $^3P^o_1$level at
267377
cm$^{-1}$  (516.3  and 259.1  10$^{-3}$ cm$^{-2}$ s$^{-1}$
respectively, see section 4.6)  and
of the line branching ratios within each of these levels, the $\ion{O}
{iii}$
$\lambda$  374.073 line
is
about six times stronger than the $\ion{O}{iii}$ $\lambda$  374.162 line
and about
3.6 times
stronger than the $\ion{O}{iii}$ $\lambda$  374.432 line. As an
illustration,  the
relative
intensities of  the five resonance lines of  $\ion{O}{iii}$ that fall
between
lambda  374.004 and lambda  374.432  are reported
in Fig. 15  together with their resulting profile for  FWHM = 100
km$^{-1}$.
From this figure one can  see that, as  a result of the strong 
contribution
of the $\ion{O}{iii}$ $\lambda$  374.073 line,  
the combined intensity near the  $\ion{N}{iii}$ $\lambda$  374.198 line   
is  higher by about 30  $\%$  than that corresponding to the $\ion{O}
{iii}$
$\lambda$  374.162
line alone. 

In this case, simple calculations of the radiative rates show that the
relative flux ratio
I$_{ 374.198}$/I$_{374.434}$ would be just about 20 $\%$ lower than
that
required (in the optically thin case) to populate the $\ion{N}{iii}$
levels 6
and 7 with the ratio required by the observations. Moreover, a
red-wing excess/asymmetry in the strong $\ion{O}{iii}$ $\lambda$  374.073
line, as
observed in all $\ion{O}{iii}$ lines, would  increase the flux
near $\lambda$  374.198 and produce a ratio close to that required.
However, we want to explicitly point out again that it is the process
of multiple  scatterings (optically thick conditions)  
which is  more effective in increasing the pumping efficiency.

\begin{figure}
\centering
\resizebox{\hsize}{!}{\includegraphics{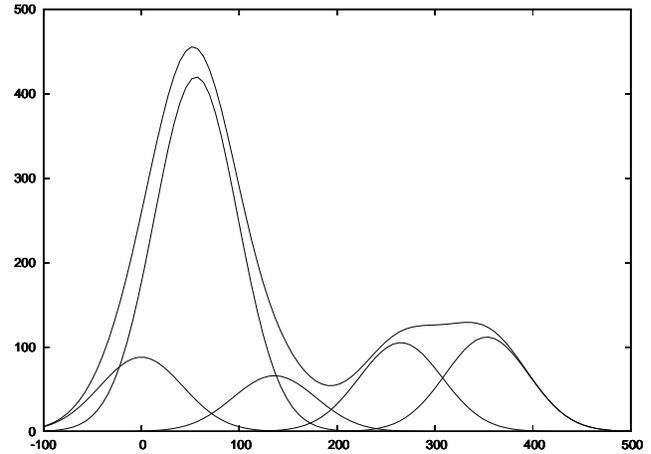}}
\caption{The  five  $\ion{O}{iii}$ resonance  lines  close to lambda
374, with
relative
height proportional to their  intensity, and  FWHM = 100 km s$^{-1}$.
The upper line represents  the sum of the five components. The x-axis
gives
the line separation in km s$^{-1}$, with zero-point at $\lambda$
 374.00.}
\end{figure}

\begin{table}
\caption{  Wavelengths (\AA) and  A$_{ij}$  values,  
for the $\ion{O}{iii}$ resonance lines  
associated with  the  Bowen fluorescence.}
\begin{center}
\renewcommand{\tabcolsep}{0.4cm}
\begin{tabular}{cccc}
\hline
 $\lambda_{vac}$ & A$_{ij}$ & $\lambda_{vac}$ & A$_{ij}$\\                              
303.413 & 3.86E+09 & 373.803 & 9.92E+08\\
303.461 & 1.16E+10 &  374.004 & 1.32E+09\\
303.517 & 2.89E+09 & 374.073 & 2.97E+09\\
303.622 & 2.89E+09 &  374.162 & 9.90E+08 \\ 
 303.695 & 4.81E+09 & 374.328 & 3.95E+09\\ 
 303.800 & 8.65E+09 &  374.432 & 1.65E+09\\
\hline
\end{tabular}
\end{center}
\end{table}
As a circumstantial support of this "new" pumping channel of level 6 of
$\ion{N}{iii}$ by $\ion{O}{iii}$ $\lambda$  374.073, despite the
separation of about 100 km
s$^{-1}$ between the
line centers, one notes that in the case of the well observed
$\ion{O}{iii}$ O3 fluorescence channel the line separation between
$\ion{He}{ii}$
Ly-$\alpha$ $\lambda$ 303.782 and  $\ion{O}{iii}$ 303.893  is not much
smaller
(about -88.2
km$^{-1}$),  while in the process that excites  the 2p3p
$^3P^o_0$ level of $\ion{O}{iii}$ and  produces the two weak lines
observed at
 $\lambda$ 3115.68 and $\lambda$ 3408.12, the
velocity separation
between the $\ion{He}{ii}$ Ly-$\alpha$ line and the $\ion{O}{iii}$
$\lambda$
303.461  line is  -250
km$^{-1}$. This latter excitation process, that is called "other" in
KB96,
necessarily requires wide wings in the resonance line of $\ion{He}{ii}$
303.782,
while the observed FWHM of the $\ion{He}{ii}$ recombination lines is
much lower,
close to 60 km$^{-1}$. This clearly indicates  that  the resonance lines
 are much wider than
the  subordinate lines.

In conclusion, the $\lambda$  374.073 resonance line of $\ion{O}{iii}$
may  also contribute to the population of level 6 of $\ion{N}{iii}$ and
thus 
increase the population ratio N$_6$/N$_7$ to a value close to the
observed  one. Obviously, similar considerations could be applied to the
$\ion{O}{iii}$ $\lambda$
 374.328
resonance
line, whose wavelength falls midway between  $\ion{N}{iii}$ $\lambda$
 374.442   and  
$\ion{N}{iii}$ $\lambda$  374.198 line and therefore would provide a
similar
contribution to the population of levels 6 and 7 of $\ion{N}{iii}$.

\subsection{ The absolute intensities of the $\ion{N}{iii}$ lines and
the
requirement of  large optical depths }

In  sect. 6.4 large optical depths in the resonance lines of O
III and $\ion{N}{iii}$ have been proposed as a mechanism  to explain the
observed intensity {\em ratio}  in the three $\ion{N}{iii}$ $\lambda$
4640
lines. This
assumption was justified by the estimates reported in the Appendix.
An additional and convincing argument in favour of this mechanism comes
from the fact  that the observed {\em absolute} intensities of
the $\ion{N}{iii}$
$\lambda$ 4640 lines necessarily require large optical depths in the
exciting lines.

We recall that from STIS and UVES observations one can directly derive
the number of $\ion{O}{iii}$ photons  that 
populate the three levels of the $\ion{O}{iii}$ 2p3s $^3P^o_J$ term from
which
the six $\ion{O}{iii}$ lines near $\lambda$ 374 have origin.
Therefore,
starting from the observed number of $\ion{O}{iii}$ photons (0.26
photons
cm$^{-2}$ s$^{-1}$)  into level
2p3s $^3P^o_1$ of $\ion{O}{iii}$ (the upper level of the $\ion{O}{iii}$
$\lambda$ 374.434
line, that can populate level 7 of $\ion{N}{iii}$) one can provide an
upper
limit for the expected number of photons
in the $\ion{N}{iii}$ $\lambda$ 4640.64 line.

 The branching ratio
from level 2p3s $^3P^o_1$ of $\ion{O}{iii}$
toward the $\lambda$  374.432 line (there are two  competing decays
from the
same level at $\lambda$  374.162 and $\lambda$  374.004) is close to
0.42
(see Fig.6 and Table 8) and therefore
about 0.11 photons cm$^{-2}$ s$^{-1}$ are expected in the $\ion{O}
{iii}$
$\lambda$  374.432 line. Assuming full
conversion  of this line into the two $\ion{N}{iii}$ lines at
$\lambda$  374.434
and  $\lambda$ 
 374.442, from the  A$_{ij}$ ratio (close to 6.0) of these two $\ion{N}
{iii}$
transitions  one 
obtains that the number of photons cm$^{-2}$ s$^{-1}$ that populate
level 7 
(3d $^2D_{5/2}$) of $\ion{N}{iii}$ through the $\lambda$ 374.434 line is
about
0.093. 
Assuming that  level 7 (3d $^2D_{5/2}$) of  $\ion{N}{iii}$  will
decay only  through
the $\lambda$ 4640.64 line (instead of the more competitive resonant
decay)
one obtains the same value  (0.093)  for the number
of photons  in the $\lambda$ 4640.64 line.
This corresponds to a
line intensity of 3.96 10$^{-13}$ erg cm$^{-2}$ s$^{-1}$.

The observed intensity (STIS) of the $\lambda$ 4640.64 line is 3.46
10$^{-13}$ 
erg
 cm$^{-2}$ s$^{-1}$ (corresponding to 0.081 photons cm$^{-2}$ s$^{-1}$).
Therefore the observed intensity of $\ion{N}{iii}$ $\lambda$ 4640.64
is only slightly less ($\sim$87$\%$) than that expected in the limiting
case of
full
conversion  of $\ion{O}{iii}$ $\lambda$   374.432 into $\ion{N}{iii}$
$\lambda$  374.434 and of $\ion{N}{iii}$ $\lambda$  374.434 into
$\ion{N}{iii}$ 
$\lambda$  4640.64 (and additional decays).

As already mentioned, in the optically thin case the conversion of
$\ion{O}{iii}$ $\lambda$  374.432 into $\ion{N}{iii}$ 374.434 by line
overlap would be
much smaller than 1,  but above all, the A$_{ij}$ branching
ratio from level 7 of $\ion{N}{iii}$
would
favour   re-emission of $\lambda$ 374.434 instead of $\lambda$  4640.64,
by a factor
(1.26 10$^{10}$)/(7.17 10$^7$) = 175  and therefore the expected
emission
intensity in the $\lambda$ 4640.64 line would be lower by a factor close
to 10$^3$. Similar considerations can be used for level 6 and the
$\ion{N}{iii}$ 
$\lambda$ 4641.85 and $\lambda$  4634.13 lines.

In conclusion,  very low $\ion{N}{iii}$ intensities are expected in the 
optically thin regime
because the A$_{ij}$ factors strongly favour re-emission of the
resonance
lines relative to the decays into the  subordinate lines.
In fact, the whole Bowen process starting from  $\ion{He}{ii}$
Ly-$\alpha$ and  the $\ion{O}{iii}$ transitions near $\lambda$  303.80 
up to the 
the $\ion{N}{iii}$ $\lambda$ 4640  lines  (via the $\ion{O}{iii}$ and 
$\ion{N}{iii}$  $\lambda$ 374
lines)  requires   high
optical depths  in the resonance lines.
In the  optically thin case the intensity of the $\ion{O}{iii}$  Bowen
lines
would be also negligible, and "a fortiori" that of the $\ion{N}{iii}$
$\lambda$
4640
lines.

These considerations strongly support our previous arguments about the 
large optical depths in the resonance lines. Thus, the requirement of
large optical depths is not an "ad-hoc" assumption but emerges in
a self-consistent  picture.

We recall that the efficiency of the $\ion{N}{iii}$ Bowen fluorescence
can be
defined (Eastman and MacAlpine, 1985) as the fraction of the $\ion{O}
{iii}$
$\lambda$  374 resonance photons arising from level 2p3s $^3P^o_1$ 
which undergoes fluorescence with $\ion{N}{iii}$ and is converted to the 
$\ion{N}{iii}$ Bowen
lines near  $\lambda$ 4640. 
In the case of RR Tel, a direct  counting of the photons gives a 
fraction  
0.134/0.259 $\sim$ 0.52. We note, however, that unlike the case of
$\ion{O}{iii}$
fluorescence in which the pumping line is unambiguously  $\ion{He}{ii}$
Ly-$\alpha$  $\lambda$  303.782, in the case of $\ion{N}{iii}$ the
candidate
pumping lines can be as high as six (depending on the line width).
Therefore the efficiency can vary from 15.3 $\%$,  if one considers all
the photons of  the six $\ion{O}{iii}$ lines near $\lambda$ 374,   to
87.0 $\%$, 
if one considers only the photons from the $\ion{O}{iii}$  
$\lambda$  374.432 line.
Eastman and MacAlpine (1985) have provided a direct expression  for the
$\ion{N}{iii}$ Bowen efficiency, based on the observed intensities of
the $\ion{O}{iii}$
and $\ion{N}{iii}$ lines.  

\begin{displaymath} y_{\ion{N}{iii}} \sim 8.7 ~~\frac{I(\ion{N}{iii}~
\lambda\lambda
4634,
4641, 4642) }{I(\ion{O}{iii}~ \lambda 3133)}\end{displaymath}

In the case of RR Tel  we obtain y$_{\ion{N}{iii}}$ $\sim$ 48 $\%$, in
good
agreement
with the  estimate from  the photon fractions.

As a final remark to this section, we point out the analogy
 in  the  $\ion{He}{ii}$ - $\ion{O}{iii}$  fluorescence 
and  in the  $\ion{O}{iii}$ - $\ion{N}{iii}$  
fluorescence: in the former case  the pumping line is the optically
thick $\ion{He}{ii}$ Ly-$\alpha$ line  while the pumped lines are the
optically
thick
resonance
lines of $\ion{O}{iii}$ at $\lambda$  303.800 (O1) and $\lambda$ 
303.695 
(O3),
that are
converted into longer wavelength subordinate lines, i.e. the primary and
secondary Bowen decays that easily escape from the nebula (the tertiary
decay is into the
six resonance lines near $\lambda$ 374). In the $\ion{O}{iii}$ -
$\ion{N}{iii}$ fluorescence
the pumping lines are the optically
thick $\ion{O}{iii}$ lines near  $\lambda$ 374 while 
the pumped lines
are the three optically thick lines of $\ion{N}{iii}$ near $\lambda$ 374
that are
converted into the subordinate $\lambda$ 4640 primary lines and the
$\lambda$ 4100 secondary lines, that also easily escape from the nebula.

\subsection{A detailed examination of the energy  balance in the
$\ion{N}{iii}$
levels }

\begin{enumerate}
\item Levels 7 (3d $^2D_{5/2}$, 267244.0
cm$^{-1}$) and  6 (3d $^2D_{3/2}$, 267238.40 cm$^{-1}$):  

these are the upper levels of the $\ion{N}{iii}$ $\lambda$ 4640.64 and
$\lambda$
 4634.13 + $\lambda$ 4641.85 lines respectively. We have searched in
"The Atomic Line List v.2.05" for all the strongest decays into these
levels
from all higher levels and checked for the possible presence of these
lines in the
spectrum of RR Tel.  This search has given a definite negative result,
primarily  for  the absence of the two very strong lines of mult. UV 24
 $\lambda$ 1885.06 and $\lambda$ 1885.22 and of other strong lines.
Therefore,    
there are no contributions to these levels  other than that of the
$\lambda$
374 lines. This 
result is in agreement with the negative result described in sect. 6.3
concerning   the absence of recombination lines.
We note that the observed
I$_{ 4634.13}$/I$_{ 4641.85}$ ratio is
 3.50, while it should be close to 5.0
irrespective of any process, because the lines share the same upper
level  and have  relative A$_{ij}$  values = 5.0. A
possible explanation is that the $\lambda$  4641.85 line is
partially  blended. Its slightly larger FWHM (Table
7) supports this suggestion.

\item level 5 (3p $^2P^o_{3/2}$, 245701.30 cm$^{-1}$):

 An examination in "The Atomic Line List 2.05"  of all
possible decays into  level 5 (3p $^2P^o_{3/2}$ (245701.30 cm$^{-1}$),
 shows that besides the
$\lambda$ 4640.64 and $\lambda$ 4641.85 lines  also other transitions
are listed  (e.g., $\lambda$
1387.38
and $\lambda$ 1805.66) with much stronger A$_{ij}$ values.   However,
these lines
are not present in the RR Tel spectrum  confirming that level 5  (3p
$^2P^o_{3/2}$)  is populated by the $\lambda$ 4640.64 and $\lambda$
4641.85 lines only, thus supporting the
fluorescence mechanism.

The $\lambda$ 4640.64 line and the $\lambda$
 4641.85 line are parents of the
$\lambda$  4097.36 line. From the
observed
intensities one can see that the sum of the number of photons 
in these two lines is larger by a factor about 2.2 than the number of
photons in the  $\lambda$  4097.36 line. Inspection of 
"The Atomic Line List 2.05"  shows that competing decays from the 3p
$^2P^o_{3/2}$ level are two EUV transitions at $\lambda$ (vac.) 691.19
and $\lambda$ 871.86.
The sum of the A$_{ij}$  in these two lines is about 1.4 times larger
than the A$_{ij}$ of the $\lambda$  4097.36 line. Therefore, 
the difference in the number of decays is almost reconciled.

\item level 4 (3p $^2P^o_{1/2}$,  245665.40 cm$^{-1}$):

This level is fed by the  $\lambda$  4634.13 line and is the upper level
of the 
$\lambda$   4103.39 line. The emission intensity of this line is not
accurately determined because it  falls in the red wing of the strong 
H$_{\delta}$ line. 

From an examination in "The Atomic Line List 2.05"  of all
possible decays into  level 4 (3p $^2P^o_{1/2}$ (245665.40 cm$^{-1}$),
 one notes that also other transitions are present (e.g., $\lambda$
1387.30
and $\lambda$ 1804.49) with much stronger A$_{ij}$ values  than those of
the
$\lambda$ 4640.64 and $\lambda$ 4641.85 lines. However, these lines
are not present in the RR Tel spectrum  confirming that level 4   is
populated by $\lambda$  4634.13 only.

From level 4  there are two competing decays at
$\lambda$ 691.40 and $\lambda$ 872.13 and the sum of the A$_{ij}$ of
these two lines 
($\sim$ 1.37 10$^8$) is about 1.6 times larger than the A$_{ij}$ ($\sim$ 
8.62 10$^7$)  of
the $\lambda$ 4103 line. The  ratio of the number of photons in the  
$\lambda$  4634.13   and $\lambda$   4103.39 lines (as obtained from
Table 7) is about 1.8 in fair agreement with the
expectations, if one takes into account the uncertainty in the emission
intensity of the $\lambda$  4634.13 line. 
\end{enumerate}

\subsection{The large optical depths in the $\ion{O}{i}$ resonance
lines}

\begin{figure}
\centering
\resizebox{\hsize}{!}{\includegraphics{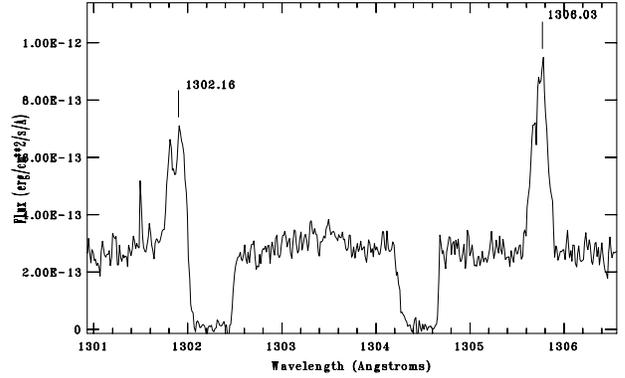}}
\caption{The resonance  $\ion{O}{i}$  lines  close to $\lambda$ 1304.
The two 
absorption features are IS and due to the zero-volt lines  of 
$\ion{O}{i}$
$\lambda$
1302.16 and Si II $\lambda$ 1304.37. This latter line has fully absorbed
the
$\ion{O}{i}$ $\lambda$ 1304.86 line.}
\end{figure}

\begin{figure}
\centering
\resizebox{\hsize}{!}{\includegraphics{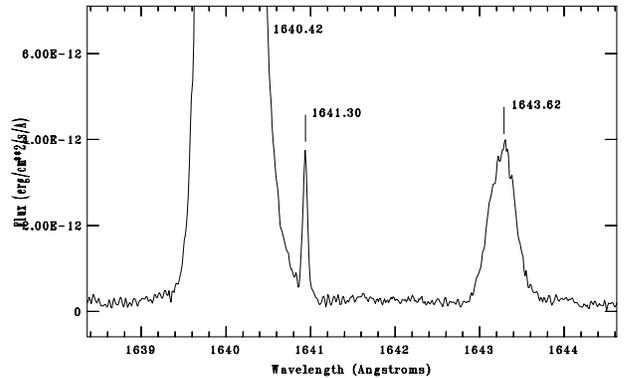}}
\caption{The $\ion{O}{i}$] $\lambda$ 1641.30 intercombination line of
mult. 146 UV
(3s $^3S^o_1$
$\to$ 
2p$^4$ $^1D_2$) on the red wing of the strong $\ion{He}{ii}$ Ba-$\alpha$
$\lambda$
1640.40  line.  The line at $\lambda_{obs}$=1643.25 is $\ion{O}{v}$
$\lambda$
1643.62. The  $\lambda$ 1641.30 line 
shares its upper level with the three $\ion{O}{i}$ resonance  lines 
close to
$\lambda$
1304}
\end{figure}

An additional argument in support of large
optical thickness in  the resonance lines  comes
from
the intensities of the three $\ion{O}{i}$ resonance lines of mult. UV 2 
near  $\lambda$ 1304 ( 2p$^3$ 3s $^3S^o_1$ $\to$  2p$^4$  $^3P_{0,1,2}$)
relative to the intercombination line at  $\lambda$  1641.30  line (3s
$^3S^o_1$
$\to$ 
2p$^4$ $^1D_2$, mult. UV 146)  with whom they share the same upper level
3s $^3S^o_1$  (76794.98 cm$^{-1}$).   
We note that there is also the  $\lambda$  2325.45 intercombination line  
(3s $^3S^o_1$
$\to$
2p$^4$
$^1S_{0}$)  that decays from the same level but its  wavelength is
nearly coincident with that of the  CII line $\lambda$  2325.40.

The emission intensity of  $\lambda$ 1302.17 (l.l. = 0.00 eV) is
affected by
partial absorption by its  IS counterpart, while the       
$\ion{O}{i}$ $\lambda$ 1304.86  line (l.l. = 158.26 cm$^{-1}$) is fully
absorbed 
by
the Si II(3)   $\lambda$ 1304.37  line  (l.l. = 0.00 eV).  Instead,
the $\lambda$  1306.03  line (l.l. 226.98 cm$^{-1}$) is  
free from IS  absorption (see Fig. 16). 
The A$_{ij}$ ratio  1306.03/1641.30 would favour the  $\lambda$  1306.03
resonance line  with respect to  $\lambda$ 1641.30  by a factor 0.65
10$^8$/1.83 10$^3$ $\sim$ 3.5 10$^4$. Instead, the observed intensites 
on
STIS are similar (1.25
10$^{-13}$ erg cm$^{-2}$ s$^{-1}$  and 2.24   10$^{-13}$ erg cm$^{-2}$
s$^{-1}$  respectively).
Evidently, whatever is the mechanism for population of the 3s $^3S^o_1$
level
(likely a decay from 3d $^3D^o$, after Ly-$\beta$ $\lambda$ 1025
fluorescence, through $\lambda$  11287 and   $\lambda$ 8446, since this
latter line is present in FEROS spectra)
the $\ion{O}{i}$  resonant photons will scatter many times  and there is
a very
small
but finite chance that at each resonant scattering decay from the  upper
level 3s $^3S^o_1$ will occur through the intercombination lines at 
$\lambda$ 
1641.30 and $\lambda$  2325.45.

Thus, if the scattering number is very high, processes with a
small
probability per scattering of destroying  resonance lines can become
important   (Osterbrock, 1962).
Unlike the $\lambda$ 1304 photons that are trapped  resonant
radiation,
the  $\lambda$ 
1641.30
photons,  will leave the nebula undisturbed  because of  their very low
A$_{ij}$ value. Thus, the large optical depth
of
the $\lambda$ 1304 lines has the effect of converting these resonantly
trapped photons into photons of the $\lambda$  1641.30  line.

The importance  of this  kind of process  involving the  $\ion{O}{i}$ 
$\lambda$ 
1641.30  line for the determination of the opacity in the resonance
lines 
has
been pointed out by Jordan (1967) and by Jordan  and Judge (1984) for
the chromospheres of cool
stars. Bhatia and Kastner (1995) have further developed this method  and
derived 
an explicit graph  of the  I$_{1641}$/I$_{1302}$ ratio as function of
log
$\tau_{1302}$. In the specific case of RR Tel they have used the
measurements  in IUE spectra by Penston et al. (1983) and derived log
$\tau_{1302}$ $\sim$
6.0. This estimate should actually be considered  as  an upper limit
because the IS absorption of
the $\ion{O}{i}$ $\lambda$ 1302 
line was not taken into account.  However, using our reliable 
I$_{1641}$/I$_{1306}$   ratio from the STIS data, and  "mutatis
mutandis"  (the A$_{ij}$
ratio of $\lambda$  1302.16  relative to  $\lambda$  1306.03 line is
near
5.0) we have  obtained again  that $\log$ $\tau_{1306}$ $\sim$  6.0. 

We also note that the resonance line
$\lambda$ 1306.03  is much wider (FWHM $\sim$ 43 km $s^{-1}$)
than the  intercombination $\ion{O}{i}$] line $\lambda$ 1641.30  (FWHM
$\sim$ 12 km $s^{-1}$) 
although  both lines 
arise from the same upper level (see also Fig. 16 and  Fig. 17).
This  supports our previous considerations about the  much larger width
of the  resonance lines as compared to the subordinate and
 intercombination ones.

\section { Conclusions.}

By combining STIS and UVES data we have obtained very accurate
measurements 
of the
absolute intensities for about 25  $\ion{He}{ii}$ Fowler lines  and for
almost all
 $\ion{O}{iii}$  lines produced by the
Bowen mechanism.  

A new measure of reddening (E$_{(B-V)}$ $\sim$ 0.00) has been obtained
from
the decrement of the $\ion{He}{ii}$ Fowler series  down to the region of
the  series limit  near $\lambda$ 2060. This new E$_{(B-V)}$ value, which
is
in contrast with the commonly assumed value E$_{(B-V)}$ $\sim$ 0.10,
has been confirmed by the observed STIS UV + optical continuum
distribution, by a re-analysis of the IUE low resolution data, and by
the value of the neutral H column density, obtained from the damped
profile of the IS Ly-$\alpha$ line.

The high quality of the spectral data has allowed us 
to obtain the most complete  set of measurements of the
Bowen fluorescence lines so far reported for any astronomical source,
and includes six "pure" O1 lines,  seven "pure"
O3 lines and the two very weak lines at $\lambda$ 3115.68  and $\lambda$
3408.12
both associated with pumping in the $\lambda$303.46 channel.
This is 
 also the first measurements of absolute intensity,
 for the two  lines at $\lambda$ 3115.68  and $\lambda$
3408.12,  both associated with pumping in the $\lambda$ 303.46 channel, 
and  the first detection of 
two weak  lines  belonging to the primary decay  of the O1 process, that
have been found 
near $\lambda$ 2180. 

The data
have also allowed   a  detailed study of the  decays from all levels
involved in the
Bowen mechanism  and a detailed comparison with the
predictions of the theoretical models of Froese Fischer (1994) and 
Kastner and Bhatia (1996). In general, the observed $\ion{O}{iii}$
emission
intensities,  are
in very good agreement with the predictions of the Froese Fischer (1994)
model.   The STIS and UVES data indicate an efficiency in  the O1 and O3
channels of about 20 $\%$ and 0.7 $\%$, respectively.
In the past   (1978 - 1995)  the O1 and O3 efficiency
 (obtained from a study of  IUE  high resolution 
spectra)  was  about 2.1 times  and  4.0 times higher, respectively

A detailed study of the $\ion{N}{iii}$ $\lambda$ 4640 emission lines and
of
their possible excitation mechanism has shown that, recombination and
continuum
fluorescence being ruled out, line fluorescence remains the only viable
excitation mechanism. We have pointed out the important role of multiple
scattering in the resonance lines and shown that the observed relative
ratios
{\em and} absolute intensities of the $\ion{N}{iii}$ lines can be
explained in
terms of line fluorescence by the three resonance lines of $\ion{O}{iii}$
at
$\lambda$$\lambda$  374.432, 374.162 and 374.073 under optically thick
conditions.

\appendix 
\section{The optical depths in the $\ion{O}{iii}$ and $\ion{N}{iii}$
resonance
lines}

The line absorption cross section per atom at the line center is

\noindent 
$\sigma_o$ = 8.30 10$^{-13}$ f$_{12}$ $\lambda_{o}^2$
($\Delta$$\lambda_{D}$)$^{-1}$, where $\Delta$$\lambda_{D}$ is the line
FWHM.
The optical depth at the line center is $\tau_{o}$=$\sigma_o$ R$_{neb}$
N$_x$,
where N$_x$ is the number density of absorbers.

\noindent
The radius of the  $\ion{O}{iii}$ emitting region  (O$^{+2}$) nearly
coincides with 
that of He$^+$ since O$^{+2}$  has an ionization potential 54.9 eV,
nearly the same
(54.4 eV) as He$^+$ (Osterbrock and Ferland, 2006). Ionization models
(ibidem) 
also  show that for an effective temperature of the exciting star greater
than 40,000 K 
(HN86 give 200,000 K for RR Tel) the He$^+$ zone coincides with the H$^+$
zone. 
Therefore, the relevant radius  can be obtained
from the observed emission intensity in the H$_{\beta}$ 4861 line,  if
N$_{e}$ and the
distance are known: 

\noindent
 L$_{4861}$=h$\nu_{4,2}$ $\alpha_{4861}$ N$_{e}$ N$_{H^+}$ 4/3 $\pi$
$R^3_{neb}$=4$\pi$d$^2$ F$_{4861}$
 
\noindent After substitution of the  numeric values
(h$\nu_{4,2}$ $\alpha_{4861}$=1.03 10$^{-25}$,
N$_{H^+}$=0.8N$_{e}$,  
4$\pi$d$^2$=1.445 10$^{45}$, F$_{4861}$=162.0 10$^{-13}$)
one  obtains that   R$^3$=5.40 10$^{58}$/N$_e^2$.

As an average of various N$_e$ diagnostics for $\ion{N}{iii}$ and
$\ion{O} {iii}$
we have obtained N$_e$$\sim$6$\cdot$10$^6$, in agreement with previous
estimates by HN86, and in fair agreement with the value (Ne=1.9 10$^7$)
very
recently obtained by Young et al. (2005) from a diagnostic based on the
ratio
of $\ion{Fe}{vii}$ lines. Therefore, in the case of a spherically
symmetric,
homogeneous nebula (filling factor=1) the radius of the H$^+$ region
is R$_{H^+}$$\sim$1.2 10$^{15}$ cm. HN86 instead obtained a value of 1.0
10$^{15}$
cm for the whole nebula. 
\noindent

For the $\ion{O}{iii}$ $\lambda$ 374.432 line one  has f$_{12}$= 2.08
10$^{-2}$,
and for its $\Delta$$\lambda_{D}$ we assume 0.045 10$^{-8}$ cm (a rather
conservative value that corresponds to a FWHM of 36 km s$^{-1}$, the
average value
from the subordinate $\ion{O}{iii}$ Bowen lines). 
Assuming that: 1) N$_e$=6$\cdot$10$^6$, 2)  N$_e$$\sim$1.2N$_{H}$,
 O/H$\sim$9 10$^{-4}$ (nearly solar, see Nussbaumer et al. 1988), 3)
 the dominant ionization stage of oxygen in a large fraction of the
nebular
volume is O$^{+2}$ (O$\sim$O$^{+2}$) (HN86), and 4) the 2p$^{2}$
$^3$P$_2$
level of the ground term of $\ion{O}{iii}$ is thermally populated, one
obtains
that the optical depth in the line center of the $\ion{O}{iii}$ 374.432
line is
$\tau_{o}$(374.432)$\sim$1790. Similar calculations for
 $\ion{O}{iii}$374.162 (f$_{12}$= 2.08 10$^{-2}$) and $\ion{O}{iii}$  
374.073 (f$_{12}$=6.32 10$^{-2}$) give $\tau_{o}$(374.162)$\sim$1075 and
 $\tau_{o}$(374.073)$\sim$5370, respectively.

The average number of scatterings $<$N$>$ in the nebula is closely
related to the line optical depth. For a rough estimate, we adopt the
common relation ($<$N$>$$\sim$$\tau$$_o$$\cdot$$\sqrt{\ln(\tau_o)}$)
that for the $\ion{O}{iii}$  lines gives  $<$N$_{374.432}$$>$$\sim$4900,
$<$N$_{374.162}$$>$$\sim$2840, and  $<$N$_{374.073}$$>$$\sim$15700.

For the $\ion{N}{iii}$ 374 lines, "mutatis mutandis", i.e.  1) N/H $\sim$
3 10$^{-4}$, 2)  $N^{+2}$ $\sim$ 0.2 N since most nitrogen is in the
N$^{+3}$
ionization stage (HN86), and 3) adopting the specific f$_{12}$ values
and
populations of the ground term,  one obtains
$\tau_{o}$(374.442)$\sim$550,
$\tau_{o}$(374.434)$\sim$1130, $\tau_{o}$( 374.198)$\sim$622, and
correspondingly that $<$N$_{374.442}$$>$$\sim$1380,
$<$N$_{374.434}$$>$$\sim$3000,
 and $<$N$_{374.198}$$>$$\sim$1580.

\end{document}